\documentclass[aps,prc,twocolumn,showpacs,superscriptaddress,floatfix]{revtex4-2}
\usepackage{graphicx}
\usepackage{dcolumn}
\usepackage{bm}
\usepackage{hyperref}
\usepackage[utf8]{inputenc}
\hypersetup{
  colorlinks=true,        
  linkcolor=blue,         
  citecolor=cyan,         
  urlcolor=cyan            
}
\usepackage{colortbl}
\usepackage{amsmath}
\usepackage{times}
\usepackage{times}
\usepackage{multirow}
\usepackage[normalem]{ulem}

\newcommand{\be}{\begin{equation}}
\newcommand{\ee}{\end{equation}}
\newcommand{\bea}{\begin{eqnarray}}
\newcommand{\eea}{\end{eqnarray}}

\begin{document}
\title{Spanning the full range of neutron star properties within a microscopic description}

\author{Tuhin Malik}
 \email{tm@uc.pt}
 \affiliation{CFisUC, Department of Physics, University of Coimbra, 3004-516 Coimbra, Portugal.}
 \author{Márcio Ferreira}
 \email{marcio.ferreira@uc.pt}
 \affiliation{CFisUC, Department of Physics, University of Coimbra, 3004-516 Coimbra, Portugal.}
\author{Milena Bastos Albino}
 \email{milena.albino@usp.br}
 \affiliation{CFisUC, Department of Physics, University of Coimbra, 3004-516 Coimbra, Portugal.}
 \author{Constan\c ca Provid\^encia}
 \email{cp@uc.pt}
 \affiliation{CFisUC, Department of Physics, University of Coimbra, 3004-516 Coimbra, Portugal.}

\date{\today}

\begin{abstract} 
The high density behavior of nuclear matter is analyzed within a relativistic mean field description with non-linear meson interactions. To assess  the model parameters and their output, a Bayesian inference technique is used. The Bayesian setup is limited only by a few nuclear saturation properties, the neutron star maximum mass larger than 2 M$_\odot$, and the low-density pure neutron matter equation of state (EOS) produced by an accurate N$^3$LO calculation in chiral effective field theory. Depending on the strength of the non-linear  scalar vector  field contribution, we have found three distinct classes of EOSs, each one correlated to different star properties distributions. If  the non-linear vector  field contribution is absent, the gravitational maximum mass and the sound velocity at high densities are the greatest. However, it also gives the smallest speed of sound at  densities below three times saturation density. On the other hand,  models with the strongest  non-linear vector  field contribution  predict the largest radii and tidal deformabilities for 1.4 M$_\odot$ stars, together with  the smallest mass for the onset of the nucleonic direct Urca processes and the smallest central baryonic densities for the maximum mass configuration.  {These models have the largest speed of sound below three times saturation density, but the smallest at high densities, in particular, above four times saturation density the speed of sound decreases approaching approximately $\sqrt{0.4}c$ at the center of the maximum mass star. On the contrary, a weak non-linear vector contribution gives a monotonically increasing speed of sound.} {A 2.75 M$_\odot$ NS maximum mass was obtained in the tail of the posterior with a weak non-linear vector field interaction. This indicates that the secondary object in GW190814 could also be an NS. {The possible onset of hyperons  and the compatibility of the different sets of models with pQCD are discussed. It is shown that pQCD favors models with a large contribution from the non-linear vector  field term or which include hyperons.}}

\end{abstract}

\maketitle

\section{Introduction}
It has been shown that the very large neutron-proton asymmetry and baryonic density that exist in the universe inside compact objects such as neutron stars (NSs), can be studied using multi-messenger astronomy, which provides us with comprehensive information far beyond what is available in terrestrial laboratories \cite{book.Haensel2007,Lattimer:2000nx,Rezzolla:2018jee}. NSs are believed to contain extremely rare phases of matter within the cores \cite{book.Glendenning1996,Burgio:2021vgk}. Using astrophysical observations together with theoretical models of the {equation of state (EOS)}, the astrophysics community is trying to understand not only the permissible domain of the EOS but also the possible scenarios of particle species pertaining to NS matter. In the case of high density matter, there is the possibility that a wide variety of phases or compositions occur,
{including  hyperons, quarks, superconducting matter, or colored superconducting matter \cite{book.Glendenning1996}}. However, up to this point in time, we know very little about NS's composition. 
{The particle composition  derived from NS matter is largely model-dependent in nature. With the present different types of available EOS models, the constraints from {the Neutron star Interior Composition Explorer (NICER) observatory}  and {gravitational waves (GW)} are still compatible with the sole inclusion of nucleonic degrees of freedom \cite{Malik:2022jqc}.}
It is imperative to note that the calculation of the nuclear EOS is a problem of theoretical modeling of the nuclear interaction. There are different models that can be used to describe the nuclear EOS of NS matter. In spite of this, relativistic mean field (RMF) models are preferred because they are capable of describing matter with relativistic effects, {important for dense matter} such as matter in NS, as well as finite nuclei \cite{Glendenning:1991es,Serot1984,Mueller:1996pm,Lalazissis:1996rd,Horowitz:2000xj,Dhiman:2007ck,Agrawal:2010wg,Chen:2014sca,Pais:2016xiu}. 

To account for the many-body effects associated with nuclear interactions, it has been established that RMF models provide a suitable description of finite nuclei and infinite nuclear matter as a result of meson exchange. A relativistic mean field model  is  built from  an effective Lorentz scalar Lagrangian that incorporates  baryon,  scalar,  and  vector meson fields \cite{Walecka:1974qa, Boguta1977, book.Glendenning1996}. The  mesonic fields  are introduced to describe the nuclear interaction: the $\sigma$ mesons generate an attractive force, while  the $\omega$ mesons generate a repulsive short-range force. Within the RMF formalism, two approaches are available to adequately describe the density dependence of the EOS and the symmetry energy. In  one of the approaches, nonlinear meson terms have been incorporated into the Lagrangian density \cite{Boguta1977, Mueller:1996pm, Horowitz:2000xj, Agrawal:2010wg, Todd-Rutel2005} while in the other approach, density-dependent coupling parameters are used to describe the nonlinearities \cite{Typel1999, Lalazissis2005, Typel2009, Malik:2022jqc},  avoiding the introduction of various nonlinear meson interaction terms. In the Lagrangian density, the coupling parameters are not completely free but are adjusted to reproduce a few well-known experimental and empirical nuclear saturation properties. To date, it is only loosely known which properties of nuclear matter  govern the high-density behavior ($\rho >>\rho_0$) \cite{Zhang:2018vrx}, but hopefully, astrophysical observations will constrain them.

{The Bayesian approach is commonly used to optimize a set of model parameters given a set of observational/theoretical constraints \cite{Imam:2021dbe,Malik:2022ilb,Coughlin:2019kqf,Wesolowski:2015fqa,Furnstahl:2015rha,Ashton2019,Landry:2020vaw,Huang:2023grj,Patra:2022yqc}.} 
In nuclear physics and  astrophysics, this method becomes a valuable tool, because it is able to determine joint posterior distributions and correlations between model parameters for a given set of fit data. Generally, Bayesian analysis of a model provides a whole snapshot of the model under the given fit data. As previously discussed, the RMF model describes dense matter EOS related to NS successfully, with density-dependent couplings or  including a few different non-linear self or cross-mesonic intersections. In light of the current observations of NS as well as  pure neutron matter constraints obtained from chiral effective field theory calculations at low densities, it is imperative to study the effects of those interactions statistically. Our previous study explored the RMF model with density-dependent couplings within a Bayesian framework \cite{Malik:2022jqc}. {This study systematically examines the RMF model with constant couplings and nonlinear mesonic interactions within a Bayesian framework.} 
In \cite{Traversi:2020aaa}, the nonlinear meson interactions in a RMF model were investigated using a Bayesian framework based solely on astrophysical data. Pure neutron matter constraints from  chiral effective field theory calculations at low densities were ignored. Indeed, low-density bounds on {pure neutron matter (PNM)} EOS from $\chi$EFT are a very strict constraint for this family of RMF models as it will be shown in the present study. Besides, higher-order interactions of $\omega$ meson (e.g., $\omega^4$) and cross interactions between the two mesons $\varrho$ and $\omega$ were not included in that study, which was restricted to the non-linear $\sigma$-meson terms introduced in \cite{Boguta1977}. {Recently, the model we will discuss in the present study has been applied to analyze the  correlations existing among  nuclear matter parameters at saturation and  neutron star properties \cite{Pradhan:2022txg}. In particular, the role the $\omega^4$ term  plays in these correlations and in controlling the maximum star mass was discussed. It was shown that the correlations are dependent on the strength of the $\omega^4$ term.  {The same model is also considered in \cite{Huang:2023grj}, where the authors take a different approach to the one of  the present study and explore the constraining power of  the astrophysical observations coming from  all the current observation (X-ray, radio, and Gravitational detection) and from simulated future X-ray missions.} }

The present study aims at analyzing a large set of parameters of  RMF models with several nonlinear meson interactions, by employing a Bayesian approach based on a given {minimal} set of fit data, in order to perform a detailed statistical analysis.  The fit data include a few nuclear saturation properties, the observation of two solar mass NS, and an estimation of the EOS of {PNM} from a $\chi$EFT calculation. Furthermore, the consistency of the obtained EOSs from marginalized posterior distributions of the model parameters   with recent measurements of the NS mass-radius by NICER and  the dimensionless tidal deformability from GW170817 by {LIGO-Virgo collaboration} will be {analized}. {In particular, we will focus our study on the high density behavior of the speed of sound. It has been shown that conditioning  the EOS built within a physics-agnostic approach to perturbative QCD calculations at high densities has a direct influence on the behavior of the speed of sound, which shows a maximum around three times saturation density or an energy density $\approx 500$~MeV fm$^{-3}$ \cite{Somasundaram:2022ztm,Altiparmak:2022bke,Gorda:2022jvk}. On the contrary, imposing just astrophysical constraints this behavior does not occur \cite{Somasundaram:2022ztm,Gorda:2022jvk}.}

The article's structure is as follows. Section \ref{formalism} introduces a brief overview of the field theoretical RMF model for the EOS at zero temperature, while Section \ref{bayes} discusses the Bayesian parameter estimation. The results of our analysis are discussed in Section \ref{results}. 
{The effect of hyperon and perturbative QCD (pQCD) constraints on the present model are discussed in Section \ref{hypqcd}.} In Section \ref{con}, the summary and conclusions are presented.

\section{Equation of state \label{formalism}}
In the present study, we consider several sets of EOSs calculated within a {RMF} description of nuclear matter based on a field theoretical approach that includes non-linear meson terms, both self-interactions and mixed terms. These non-linear terms are important to define the density dependence of the {EOS}.  Different regions of the parameter space that give an equally good description of the nuclear properties will be considered.
The nuclear interaction between nucleons is introduced through the exchange of  the scalar-isoscalar meson $\sigma$, the vector-isoscalar meson $\omega$ and the vector-isovector meson $\varrho$.
 The Lagrangian {describing the baryonic degrees of freedom} is given by
\begin{equation}
  \mathcal{L}=   \mathcal{L}_N+ \mathcal{L}_M+ \mathcal{L}_{NL}
\end{equation} 
with
\begin{equation}
\begin{aligned}
\mathcal{L}_{N}=& \bar{\Psi}\Big[\gamma^{\mu}\left(i \partial_{\mu}-g_{\omega} \omega_{\mu}-
g_{\varrho} {\boldsymbol{t}} \cdot \boldsymbol{\varrho}_{\mu}\right) \\
&-\left(m-g_{\sigma} \phi\right)\Big] \Psi \\
\mathcal{L}_{M}=& \frac{1}{2}\left[\partial_{\mu} \phi \partial^{\mu} \phi-m_{\sigma}^{2} \phi^{2} \right] \\
&-\frac{1}{4} F_{\mu \nu}^{(\omega)} F^{(\omega) \mu \nu} 
+\frac{1}{2}m_{\omega}^{2} \omega_{\mu} \omega^{\mu} \nonumber\\
&-\frac{1}{4} \boldsymbol{F}_{\mu \nu}^{(\varrho)} \cdot \boldsymbol{F}^{(\varrho) \mu \nu} 
+ \frac{1}{2} m_{\varrho}^{2} \boldsymbol{\varrho}_{\mu} \cdot \boldsymbol{\varrho}^{\mu}.\\
    			\mathcal{L}_{NL}=&-\frac{1}{3} b~m~ g_\sigma^3 (\sigma)^{3}-\frac{1}{4} c g_\sigma^4 (\sigma)^{4}+\frac{\xi}{4!} g_{\omega}^4 (\omega_{\mu}\omega^{\mu})^{2} \nonumber\\&+\Lambda_{\omega}g_{\varrho}^{2}\boldsymbol{\varrho}_{\mu} \cdot \boldsymbol{\varrho}^{\mu} g_{\omega}^{2}\omega_{\mu}\omega^{\mu},
\end{aligned}
\label{lagrangian}
\end{equation}
The field $\Psi$ is a Dirac spinor that describes the nucleon doublet (neutron and proton) with a  bare mass $m$; 
$\gamma^\mu $  are the Dirac matrices and $\boldsymbol{t}$ is the isospin operator. The vector meson  tensors are defined as   $F^{(\omega, \varrho)\mu \nu} = \partial^ \mu A^{(\omega, \varrho)\nu} -\partial^ \nu A^{(\omega, \varrho) \mu}$.  $g_{\sigma}$, $g_{\omega}$ and $g_{\varrho}$ are the couplings  of the nucleons to the meson fields $\sigma$, $\omega$ and $\varrho$, having masses, respectively,  $m_\sigma$, $m_\omega$ and $m_\varrho$. 

The parameters $b,\, c,$ $\xi$ and  $\Lambda_{\omega}$, which define the strength of the non-linear terms, are determined together with the couplings $g_i$ $i=\sigma,\, \omega,\,\varrho$, imposing a set of constraints.  The terms with $b,\, c,$ have been introduced in \cite{Boguta1977} to control the nuclear matter incompressibility at saturation. The $\xi$ term controls the stiffness of the high-density EOS, the larger it is the softer the EOS. The $\Lambda_{\omega}$ parameter affects the density dependence of the symmetry energy, the larger the smaller the symmetry energy slope at saturation. The effect of the nonlinear terms on the magnitude of the meson fields is clearly seen from the equations of motion for the mesons
		\begin{eqnarray}
			{\sigma}&=& \frac{g_{\sigma}}{m_{\sigma,{\rm eff}}^{2}}\sum_{i} \rho^s_i\label{sigma}\\
			{\omega} &=&\frac{g_{\omega}}{m_{\omega,{\rm eff}}^{2}} \sum_{i} \rho_i \label{omega}\\
			{\varrho} &=&\frac{g_{\varrho}}{m_{\varrho,{\rm eff}}^{2}}\sum_{i} t_{3} \rho_i, \label{rho}
		\end{eqnarray}
 where $\rho^s_i$ and $\rho_i$ are, respectively, the scalar density and the number density of nucleon $i$, and
 \begin{eqnarray}
   m_{\sigma,{\rm eff}}^{2}&=& m_{\sigma}^{2}+{ b g_\sigma^3}{\sigma}+{c g_\sigma^4}{\sigma}^{2} \label{ms} \\ 
    m_{\omega,{\rm eff}}^{2}&=& m_{\omega}^{2}+ \frac{\xi}{3!}g_{\omega}^{4}{\omega}^{2} +2\Lambda_{\omega}g_{\varrho}^{2}g_{\omega}^{2}{\varrho}^{2}\label{mw}\\
    m_{\varrho,{\rm eff}}^{2}&=&m_{\varrho}^{2}+2\Lambda_{\omega}g_{\omega}^{2}g_{\varrho}^{2}{\omega}^{2}, \label{mr}
 \end{eqnarray}
 where the meson fields should be interpreted as their expectation values. Some conclusions can be drawn from these equations with respect to the density behavior of the EOS:
a)  the effective mass of the $\omega$-meson $m_{\omega,{\rm eff}}$ increases as the $\omega$-field increases and as a result at high densities $\omega\propto \rho^\alpha $ with $\alpha<1$, giving rise to a softening of the EOS  at high densities with respect to  models with a zero or small $\xi$. This will also affect the behavior of the speed of sound as we will discuss later; b) the effective mass of the $\varrho$-meson, $m_{\varrho,{\rm eff}}$, increases with the increase of the density and, as a result, the $\varrho$ field becomes weaker, which implies a softer symmetry energy. Notice, however, that if  $\xi\ne 0$ this softening is smaller since the $\omega$ field does not grow so fast with the baryonic density.

Based on a reasonable approximation, the EOS of nuclear matter can be divided into two parts: (i) the EOS of symmetric nuclear matter (SNM) $\epsilon(\rho,0)$ (ii) a term involving the symmetry energy coefficient $S(\rho)$ and the asymmetry $\delta$,
\bea
 \epsilon(\rho,\delta)\simeq \epsilon(\rho,0)+S(\rho)\delta^2,
 \label{eq:eden}
\eea 
where  $\epsilon$ is the energy per nucleon at a given density  $\rho$  and isospin asymmetry $\delta=(\rho_n-\rho_p)/\rho$.
The EOS can be recast in terms of various properties of bulk nuclear matter of order $n$ at saturation density: (i) for the symmetric nuclear matter, the energy per nucleon $\epsilon_0=\epsilon(\rho_0,0)$ ($n=0$), the incompressibility coefficient $K_0$ ($n=2$), the
skewness  $Q_0$ ($n=3$),  and  the kurtosis $Z_0$ ($n=4$), respectively, given by
\begin{equation}
X_0^{(n)}=3^n \rho_0^n \left (\frac{\partial^n \epsilon(\rho, 0)}{\partial \rho^n}\right)_{\rho_0}, \, n=2,3,4;
\label{x0}
\end{equation}
(ii) for the symmetry energy,  the symmetry energy at saturation 
 $J_{\rm sym,0}$ ($n=0$), 
\begin{equation}
J_{\rm sym,0}= S(\rho_0)=\frac{1}{2} \left (\frac{\partial^2 \epsilon(\rho,\delta)}{\partial\delta^2}\right)_{\delta=0},
\end{equation}
the slope $L_{\rm sym,0}$ ($n=1$),  the curvature $K_{\rm sym,0}$ ($n=2$),  the skewness $Q_{\rm sym,0}$ ($n=3$), and  the kurtosis $Z_{\rm sym,0}$ ($n=4$), 
respectively, defined as
\begin{equation}
X_{\rm sym,0}^{(n)}=3^n \rho_0^n \left (\frac{\partial^n S(\rho)}{\partial \rho^n}\right )_{\rho_0},\, n=1,2,3,4.
\label{xsym}
\end{equation}

\section{The Bayesian setup \label{bayes}}
By updating a prior belief (i.e., a prior distribution) with given information (i.e., observed or fit data) and optimizing a likelihood function, a posterior distribution can be obtained according to Bayes' theorem \cite{Gelman2013}. Hence, in order to set up a Bayesian parameter optimization system, four things must be defined: the prior, the likelihood function, the fit data, and the sampler. 

{\it The prior --} First, we examine the prior domain of the adopted RMF model, which provides relatively wide nuclear matter saturation properties through Latin hypercube sampling, in order to define the prior distribution of our Bayesian setup. Finally, we determine the uniform priors for each parameter listed in Table \ref{tab2}. 
\begin{table}[]
\caption{{The uniform prior  is considered for the parameters of the RMF models. Specifically, B and C are $b \times 10^3$ and $c \times 10^3$, respectively.  The entrances  ’min’ and ’max’ denote the minimum and maximum values of the distribution.}}
\label{tab2}
\setlength{\tabcolsep}{15.5pt}
\renewcommand{\arraystretch}{1.1}
\begin{tabular}{cccc}
\toprule
\multirow{2}{*}{No} & \multirow{2}{*}{Parameters}  & \multicolumn{2}{c}{\it Set 0} \\ \cline{3-4} 
                    &                                & min  & max  \\ \hline
1                   & $g_{\sigma}$                              & 6.5  & 15.5   \\
2                   & $g_{\omega}$                              & 6.5  & 15.5   \\
3                   & $g_{\varrho}$                        & 6.5    & 16.5  \\
4                   & $B$                               & 0.5    & 9.0 \\
5                   & $C$                            & -5.0    & 5.0 \\
6                   & $\xi$                                & 0.0 & 0.04  
\footnote 
{Note: We have also performed another three identical studies but for three different ranges of a uniform prior for parameter $\xi$: i) $\xi\in[0,0.004]$ ({\it Set 1}), ii) $\xi \in [0.004,0.015]$({\it Set 2}) and iii) $\xi \in [0.015,0.04]$ ({\it Set 3} ).}\\ 
7                   & $\Lambda_\omega$                      & 0& 0.12  \\
\hline
\end{tabular}
\end{table}

{\it The fit data--} In Table \ref{tab1}, the fit data include the nuclear saturation density $\rho_0$, the binding energy per nucleon $\epsilon_0$, the incompressibility coefficient $K_0$, and the symmetry energy $J_{\rm sym,0}$,  all assessed at the nuclear saturation density $\rho_0$. Additionally, we take into account the pressure of {PNM} for densities of  0.08, 0.12, and 0.16 fm$^{-3}$ from N$^3$LO calculation in $\chi$EFT \cite{Hebeler2013}, accounting for 2 $\times$ N$^3$LO data uncertainty as well as the NS maximum mass above 2.0 M$_\odot$ {with uniform probability} in the likelihood.

{\it The Log-Likelihood--} With our setup, we have optimized a log-likelihood as a cost function. For all the data presented in Table \ref{tab1}, with the appropriate $\sigma$ uncertainty, equation \ref{loglik} shows the log-likelihood function, except for the low-density PNM data and the maximum mass of NS. Our approach has been to use the box function probability as given in equation \ref{pnmlik} for the PNM data from $\chi$EFT. We also use the step function probability for the NS mass. 
\begin{equation}
\label{loglik}
    Log (\mathcal{L}) = -0.5 \times \sum_j  \left\{ \left(\frac{d_j-m_j(\boldsymbol{\theta})}{\sigma_j}\right)^2 + Log(2 \pi \sigma_j^2)\right\} 
\end{equation} 
\begin{equation}
\label{pnmlik}
    Log (\mathcal{L}) = Log \left\{ \prod_j \frac{1}{2  \sigma_j} \frac{1}{\exp \left(\frac{|d_j-m_j(\boldsymbol{\theta})|-\sigma_j}{0.015}\right)+1} \right\}
\end{equation} 
Specifically, $j$ runs over the entire dataset and $d_j$ and $m_j$ represent the data and derived model values, respectively. $\sigma_j$ represents the uncertainty associated with each data point in the dataset and the $\boldsymbol{\theta}$ is the vector representation of the model parameter. It is important to understand that when sampling the posterior, the normalization of the log-likelihood, which is done in equations \ref{loglik} and \ref{pnmlik} is irrelevant. However, to calculate the Bayes {\it evidence} it is mandatory and in some cases, it also reduces the computation time.

{To populate the six-dimensional posterior, we use the nested sampling algorithm,} first proposed in Ref. \cite{Skilling2004} and suitable for low-dimensional problems.  The PyMultinest   sampler is invoked to generate samples for the four thousand starting "n-live" points \cite{Buchner:2014nha,buchner2021nested}. 
There are approximately eighteen thousand samples we have obtained in each posterior with $\approx 0.04$ acceptance rate.

\begin{table}[]
\centering
 \caption{The constraints imposed in the Bayesian inference to generate all sets of models: 
 binding energy per nucleon $\epsilon_0$,  incompressibility $K_0$, symmetry energy  $J_{\rm sym,0}$  at the nuclear saturation density  $\rho_0$, including an 1$\sigma$ uncertainty;  the pressure of pure neutron matter  PNM determined at the densities 0.08, 0.12 and 0.16~fm$^{-3}$ from  a  $\chi$EFT  calculation \cite{Hebeler2013}, with {2 $\times$} N$^3$LO {uncertainty}  in the likelihood, the pressure of PNM is an increasing function of density and the maximum NS mass above 2$M_\odot$.}
  \label{tab1}
 \setlength{\tabcolsep}{5.5pt}
      \renewcommand{\arraystretch}{1.1}
\begin{tabular}{cccc}
\hline 
\hline 
\multicolumn{4}{c}{Constraints}                                                        \\
\multicolumn{2}{c}{Quantity}                     & Value/Band  & Ref     \\ \hline
\multirow{3}{*}{\shortstack{NMP \\  {[}MeV{]} }} 
& $\rho_0$ & $0.153\pm0.005$ & \cite{Typel1999}    \\
& $\epsilon_0$ & $-16.1\pm0.2$ & \cite{Dutra:2014qga}   \\
                               & $K_0$           & $230\pm40$   & \cite{Shlomo2006,Todd-Rutel2005}    \\
                              & $J_{\rm sym, 0}$           & $32.5\pm1.8$  & \cite{Essick:2021ezp}   \\
                              
                               &                 &                &                                                   \\
  \shortstack{PNM \\ {[}MeV fm$^{-3}${]}}                  & $P(\rho)$       & $2\times$ N$^{3}$LO    & \cite{Hebeler2013}   \\
  &$dP/d\rho$&$>0$&\\
\shortstack{NS mass \\ {[}$M_\odot${]}}        & $M_{\rm max}$   & $>2.0$     &  \cite{Fonseca:2021wxt}      \\ 

\hline 
\end{tabular}
\end{table}

\section{Results \label{results}}
\begin{figure}
\includegraphics[width=0.99\linewidth]{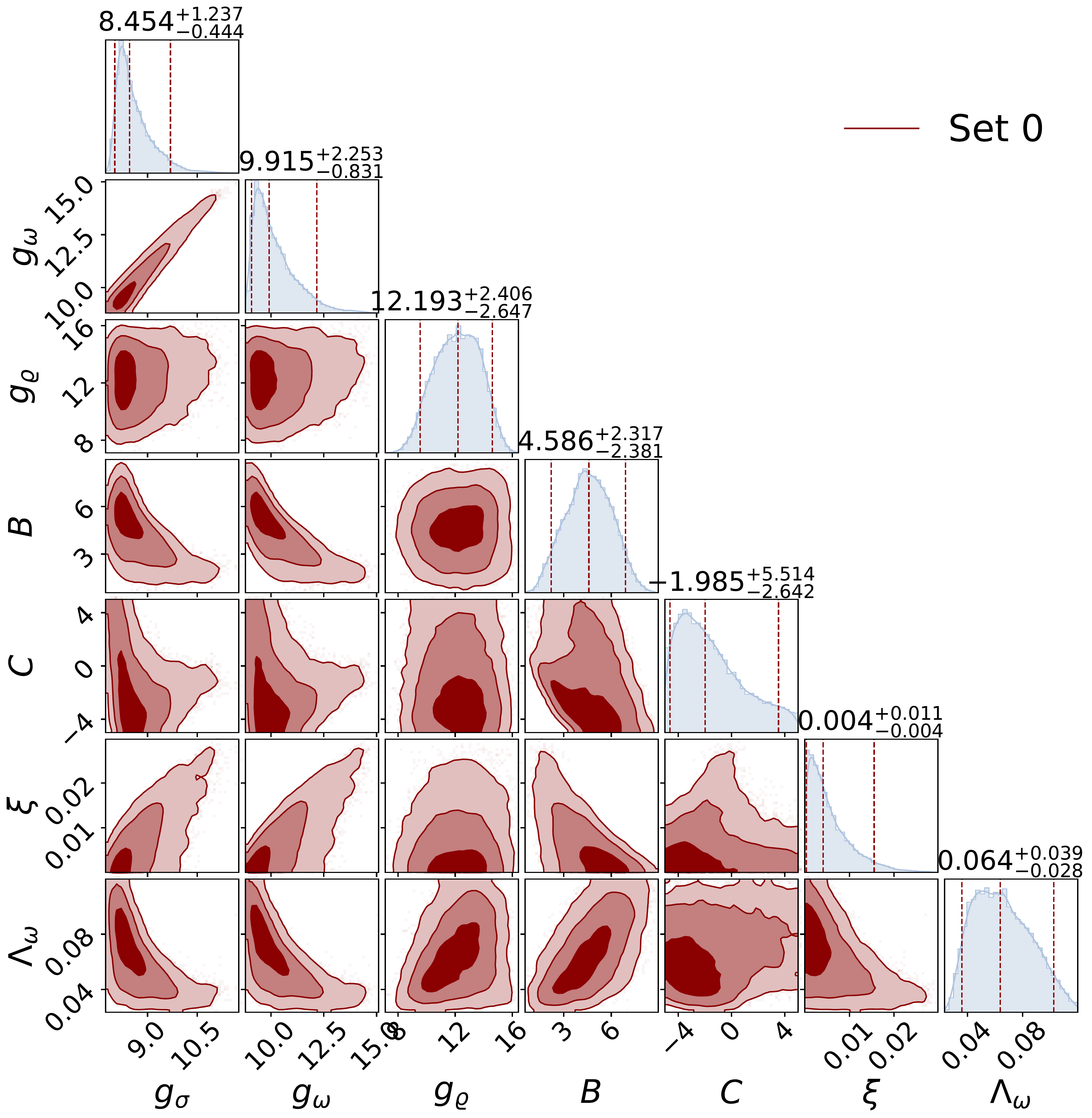}
\caption{Corner plot for the posteriors of the parameters $g_\sigma$, $g_\omega$, $g_\rho$, $B=b\times 10^3$, $C=c\times 10^3
$,  $\Lambda_\omega$,  and $\xi$ of the RMF model used in the present study obtained using the uniform priors defined in Table \ref{tab2}. The vertical lines represent the 90\% credible intervals (CIs), and the light and dark intensities represent the 1$\sigma$, 2$\sigma$, and 3$\sigma$ CIs, respectively. 
\label{T:fig2}}
\end{figure}

{In the following, we examine the posterior probability distributions of the RMF model parameters we have adapted for the purpose of this work, namely $g_\sigma, \, g_\omega,\, g_\varrho,\, b,\, c,\, \xi$, and $\Lambda_\omega$ as {briefly outlined} in Sec. \ref{bayes}. Our Bayesian setup for the RMF model parameters includes the uniform ("un-informative") prior as discussed in the earlier section.}
We first perform a Bayesian inference with prior {\it Set 0}, as given in Table \ref{tab2}, imposing the constraints given in Table \ref{tab1}. 
{Besides the conditions used in \cite{Malik:2022jqc}, the  PNM condition was implemented with hard cuts, and an extra constraint was introduced: it was imposed that the PNM pressure is an increasing function of the density.} This last condition is necessary because this behavior is physically justified but the inference process may originate models that satisfy all the other constraints except this one. In Fig. \ref{T:fig2} the corner plot for the posteriors of the parameters $g_\sigma$, $g_\omega$, $g_\rho$, $B$, $C$,  $\Lambda_\omega$  and $\xi$ is shown. The parameters $B$ and $C$ are $b \times 10^3$ and $c \times 10^3$, respectively. 

\begin{figure}
\includegraphics[width=0.99\linewidth]{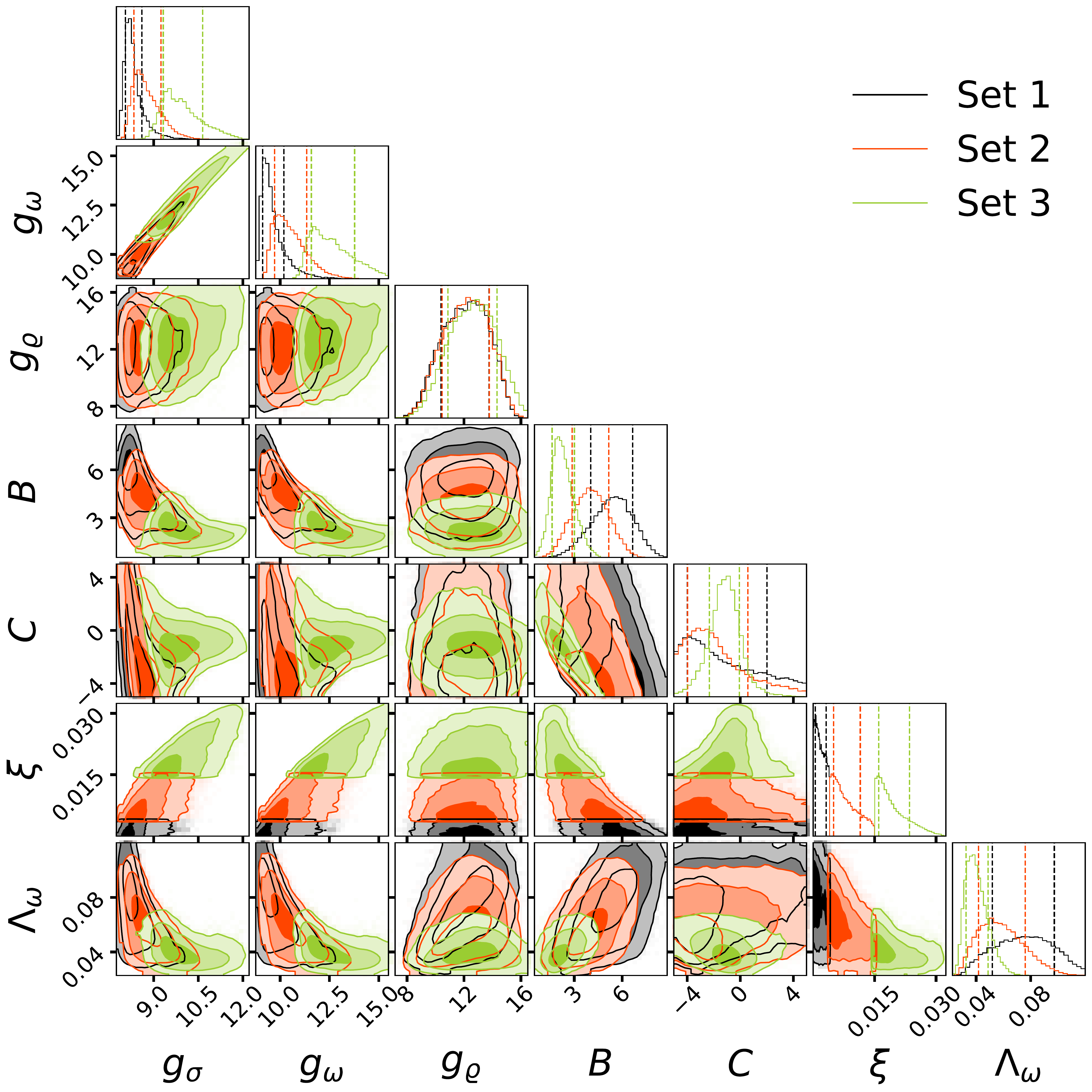}
\caption{Corner plot for the three sets of models, set 1 with $\xi\in[0,0.004]$ (solid black lines), set 2 with $\xi\in[0.004,0.015]$ (red) and set 3 with $\xi\in[0.015,0.04]$ (green), comparing the posteriors of the parameters $g_\sigma$, $g_\omega$, $g_\rho$, $B=b\times 10^3$, $C=c\times 10^3$, and $\Lambda_\omega$ of the RMF model used in the present study. The vertical lines represent the 68\% CIs, and the different intensities, from dark to light,  represent the 1$\sigma$, 2$\sigma$, and 3$\sigma$ CIs, respectively.
\label{T:fig2b}}
\end{figure}

Some comments are in order: a) {some models appear at large $g_\sigma$, $g_\omega$ and $\xi$ and small $\Lambda_\omega$.} It is the value of $\xi$ that defines this subset, and, therefore, in order to better understand the properties of these models, an independent  Bayesian inference calculation  is performed taking a {prior restriction on the parameter $\xi\in[0.015,0.04]$} ({\it Set 3}); b) in order to completely understand the effect of the $\omega^4$ term, that has a strong effect on the density dependence of the SNM EOS, in particular, determines the high-density dependence of the EOS,   two other calculations will be performed, one with $\xi\in[0,0.004]$ ({\it Set 1}) and a second with $\xi\in[0.004,0.015]$ ({\it Set 2}).

The corner plots that compare  the three sets of $\approx$ 20000 models each defined by a different constraint on $\xi$ are  shown in Figs. \ref{T:fig2b}, \ref{T:fig4} and \ref{T:fig5}, respectively,  for  the model parameters, the nuclear matter properties and NS properties  (set 1 represented by solid black lines,  set 2  by  red and set 3 by green).  The median values and associated 90\% credible intervals (CI) have been compiled in Table \ref{tab:prop}. In the table, we have listed the NMPs defined in Eqs. (\ref{x0}) and (\ref{xsym}), and the following NS properties: the gravitational mass of the maximum mass configuration $M_{\rm max}$, and corresponding  baryonic mass $M_{\rm B, max}$, radius $R_{\rm max}$, central energy density $\varepsilon_ c$, central  baryonic number density  $\rho_c$, and square of central speed-of-sound $c_s^2$ of the maximum mass NS,  the radius   $R_{{\rm M}_i}$ and  the dimensionless tidal deformability $ \Lambda_{{\rm M}_i}$  ( $\Lambda_{{\rm M}_i}$ of stars with gravitational mass ${\rm M}_i \in [1.4,1.6,1.8,2.08]$ $M_\odot$), and the effective tidal deformability $\tilde \Lambda$ for the GW170817 merger with $q=1$ ($q$ is the mass ratio of NSs engaged in the binary merger) computed for the three sets.

\begin{table*}[]
\caption{The median values and associated 90\% CI of the NMPs defined in Eqs. (\ref{x0}) and (\ref{xsym}), and NS properties, the gravitational mass $M_{\rm max}$, baryonic mass $M_{\rm B, max}$, radius $R_{\rm max}$, central energy density $\varepsilon_ c$, central number density for baryon $\rho_c$, and square of central speed-of-sound $c_s^2$ of the maximum mass NS, the radius   $R_{{\rm M}_i}$ and  the dimensionless tidal deformability $\Lambda_{{\rm M}_i}$ for NS mass ${\rm M}_i \in [1.4,1.6,1.8,2.08]$ $M_\odot$, and the effective tidal deformability $\tilde \Lambda$ for the GW170817 merger with $q=1$ ($q$ is the mass ratio of NSs engaged in the binary merger) computed for the three situations Set 1 ($\xi\in[0,0.004]$), Set 2 ($\xi \in[0.004,0.015]$), and Set 3 ($\xi \in[0.015,0.04]$) are displayed. \label{tab:prop}}
 \setlength{\tabcolsep}{5.5pt}
      \renewcommand{\arraystretch}{1.2}
\begin{tabular}{cccccccccccccc}
\toprule
\multicolumn{2}{c}{\multirow{3}{*}{Quantity}}    & \multirow{3}{*}{Units} & \multicolumn{3}{c}{Set 1}                               &  & \multicolumn{3}{c}{Set 2}                               &  & \multicolumn{3}{c}{Set 3}                               \\ \cline{4-6} \cline{8-10} \cline{12-14} 
\multicolumn{2}{c}{}                             &                        & \multirow{2}{*}{median} & \multicolumn{2}{c}{$90\%$ CI} &  & \multirow{2}{*}{median} & \multicolumn{2}{c}{$90\%$ CI} &  & \multirow{2}{*}{median} & \multicolumn{2}{c}{$90\%$ CI} \\
\multicolumn{2}{c}{}                             &                        &                         & min           & max           &  &                         & min           & max           &  &                         & min           & max           \\ \hline
\multirow{11}{*}{NMP} & $\rho_0$                 & fm$^{-3}$              & $0.152$                 & $0.145$       & $0.160$       &  & $0.152$                 & $0.145$       & $0.160$       &  & $0.153$                 & $0.145$       & $0.161$       \\
                      & $m^\star$                & …                      & $0.76$                  & $0.69$        & $0.78$        &  & $0.72$                  & $0.64$        & $0.76$        &  & $0.63$                  & $0.55$        & $0.69$        \\
                      & $\varepsilon_0$          & \multirow{9}{*}{MeV}   & $-16.10$                & $-16.43$      & $-15.76$      &  & $-16.10$                & $-16.43$      & $-15.76$      &  & $-16.10$                & $-16.43$      & $-15.77$      \\
                      & $K_0$                    &                        & $257$                   & $234$         & $293$         &  & $252$                   & $205$         & $300$         &  & $232$                   & $169$         & $295$         \\
                      & $Q_0$                    &                        & $-444$                  & $-497$        & $-301$        &  & $-438$                  & $-548$        & $-256$        &  & $-319$                  & $-562$        & $483$         \\
                      & $Z_0$                    &                        & $1766$                  & $435$         & $3054$        &  & $2161$                  & $65$          & $5521$        &  & $4698$                  & $739$         & $9623$        \\
                      & $J_{\rm sym,0}$          &                        & $31.87$                 & $29.10$       & $34.22$       &  & $31.90$                 & $29.05$       & $34.44$       &  & $32.05$                 & $29.19$       & $34.75$       \\
                      & $L_{\rm sym,0}$          &                        & $35$                    & $21$          & $57$          &  & $39$                    & $25$          & $58$          &  & $50$                    & $35$          & $64$          \\
                      & $K_{\rm sym,0}$          &                        & $-126$                  & $-177$        & $-57$         &  & $-96$                   & $-160$        & $4$           &  & $-6$                    & $-89$         & $71$          \\
                      & $Q_{\rm sym,0}$          &                        & $1438$                  & $640$         & $1736$        &  & $1328$                  & $722$         & $1661$        &  & $866$                   & $-88$         & $1303$        \\
                      & $Z_{\rm sym,0}$          &                        & $-12118$                & $-19290$      & $236$         &  & $-13057$                & $-19030$      & $-1147$       &  & $-13422$                & $-17643$      & $-6877$       \\
                      &                          &                        &                         &               &               &  &                         &               &               &  &                         &               &               \\
\multirow{15}{*}{NS}  & $M_{\rm max}$            & M $_\odot$             & $2.073$                 & $2.013$       & $2.306$       &  & $2.064$                 & $2.011$       & $2.244$       &  & $2.048$                 & $2.010$       & $2.162$       \\
                      & $M_{\rm B, max}$         & M $_\odot$             & $2.457$                 & $2.378$       & $2.772$       &  & $2.437$                 & $2.367$       & $2.677$       &  & $2.400$                 & $2.348$       & $2.546$       \\
                      & $c_{s}^2$                & $c^2$                  & $0.63$                  & $0.58$        & $0.70$        &  & $0.52$                  & $0.46$        & $0.58$        &  & $0.43$                  & $0.39$        & $0.45$        \\
                      & $\rho_c$                 & fm$^{-3}$              & $1.079$                 & $0.914$       & $1.138$       &  & $1.036$                 & $0.899$       & $1.099$       &  & $0.972$                 & $0.883$       & $1.035$       \\
                      & $\varepsilon_{c}$        & MeV fm$^{-3}$          & $1377$                  & $1169$        & $1462$        &  & $1302$                  & $1127$        & $1394$        &  & $1198$                  & $1084$        & $1288$        \\
                      & $R_{\rm max}$            & \multirow{5}{*}{km}    & $10.75$                 & $10.46$       & $11.52$       &  & $11.03$                 & $10.69$       & $11.74$       &  & $11.47$                 & $11.07$       & $11.97$       \\
                      & $R_{1.4}$                &                        & $12.34$                 & $12.03$       & $12.89$       &  & $12.50$                 & $12.17$       & $13.05$       &  & $12.87$                 & $12.42$       & $13.30$       \\
                      & $R_{1.6}$                &                        & $12.21$                 & $11.89$       & $12.86$       &  & $12.39$                 & $12.04$       & $13.02$       &  & $12.77$                 & $12.31$       & $13.26$       \\
                      & $R_{1.8}$                &                        & $11.98$                 & $11.62$       & $12.79$       &  & $12.18$                 & $11.79$       & $12.93$       &  & $12.57$                 & $12.09$       & $13.14$       \\
                      & $R_{2.075}$              &                        & $11.67$                 & $10.96$       & $12.86$       &  & $11.88$                 & $11.21$       & $12.92$       &  & $12.25$                 & $11.65$       & $12.96$       \\
                      & $\Lambda_{1.4}$          & \multirow{5}{*}{…}     & $399$                   & $338$         & $545$         &  & $439$                   & $366$         & $587$         &  & $535$                   & $420$         & $672$         \\
                      & $\Lambda_{1.6}$          &                        & $156$                   & $129$         & $233$         &  & $174$                   & $141$         & $250$         &  & $215$                   & $166$         & $284$         \\
                      & $\Lambda_{1.8}$          &                        & $62$                    & $49$          & $107$         &  & $71$                    & $55$          & $114$         &  & $89$                    & $67$          & $127$         \\
                      & $\Lambda_{2.075}$        &                        & $17$                    & $9$           & $42$          &  & $20$                    & $12$          & $43$          &  & $26$                    & $16$          & $43$          \\
                      & $\tilde \Lambda_{q=1.0}$ &                        & $474$                   & $402$         & $639$         &  & $519$                   & $434$         & $688$         &  & $631$                   & $497$         & $787$         \\ \hline
\end{tabular}
\end{table*}

First, let's discuss the model parameters for the three sets {based on the constraints on $\xi$.} The main finding is that the parameters of sets 1 and 2 do not differ much: $g_\sigma$ and $g_\omega$ extend to slightly larger values, while $B$ and $\Lambda_\omega$ take slightly smaller values. In order to compensate for the $\omega^4$ term, that softens the EOS, the $g_\omega$ must increase, a change that reflects itself on the other parameters.  
Finally, set 3 differs a lot from the other two: it spreads to larger values of $g_\sigma$ and $g_\omega$, smaller values of $B$ and $\Lambda_\omega$ and $C$ takes mainly negative values. Only $g_\rho$ is similar for the three sets. These differences will reflect on the NMP and NS properties.

\begin{figure}[!t]
\includegraphics[width=0.95\linewidth]{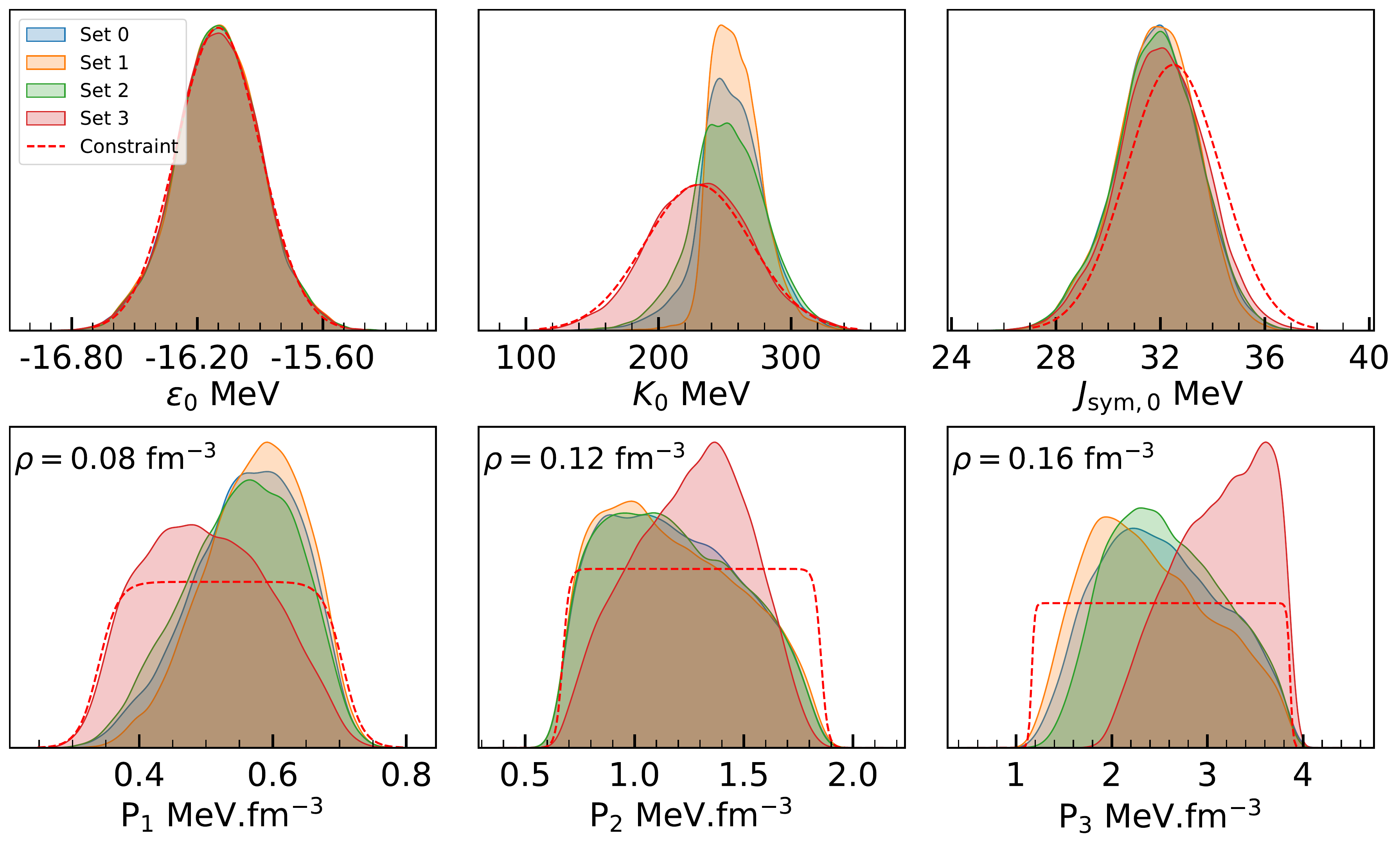}
\caption{The comparison of the marginalized  posteriors  and the corresponding 
constraints imposed in the Bayesian inference analysis.
\label{T:fig3}}
\end{figure}

\begin{figure*}[!h]
\includegraphics[width=0.8\linewidth]{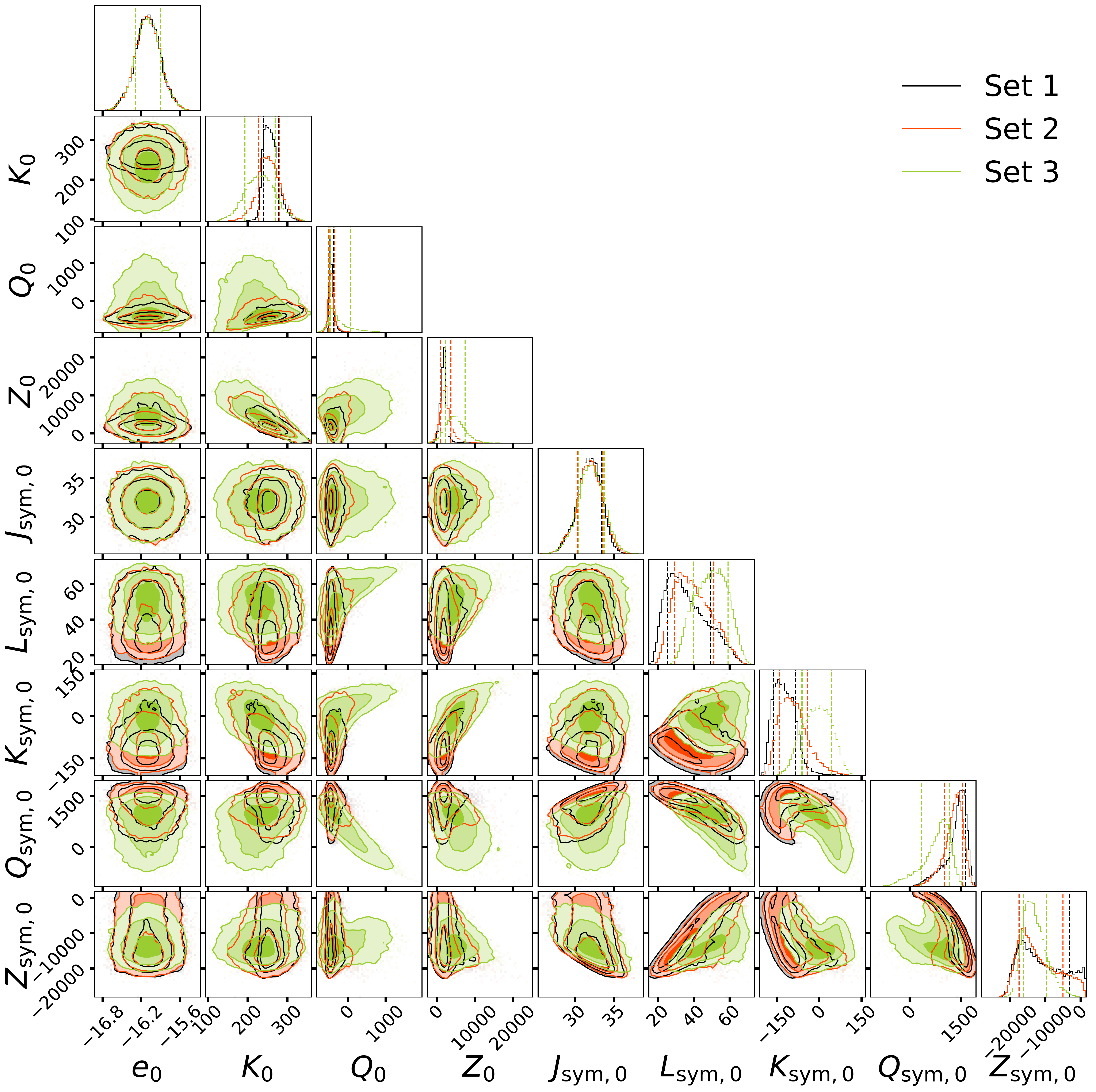}
\caption{Corner plot for the three sets of models with $\xi\in[0,0.004]$ (solid black lines), $\xi\in[0.004,0.015]$ (red) and $\xi\in[0.015,0.04]$ (green) comparing the respective nuclear matter  properties, in particular, the binding energy $e_0$, incompressibility $K_0$, skewness $Q_0$ and curtosis $Z_0$ at saturation that characterizes symmetric nuclear matter and symmetry energy $J_{\rm sym,0}$, its slope $L_{\rm sym,0}$, curvature $K_{\rm sym,0}$, skewness $Q_{\rm sym,0}$ and curtosis $Z_{\rm sym,0}$ at saturation that characterizes the symmetry energy. The vertical lines represent the 68\% CIs, and the light and dark intensities represent the 1$\sigma$, 2$\sigma$, and 3$\sigma$ CIs, respectively.\label{T:fig4}}
\end{figure*}

It is also interesting to discuss how efficiently do the posterior distributions of the nuclear matter properties specified in Table \ref{tab1} span the target distributions. 
In Fig. \ref{T:fig3}, the distributions of the posteriors of the physical properties that define the constraints imposed in the Bayesian inference given in Table \ref{tab1}  are compared with the target distributions. We conclude that: a) set 1 and 2 have very similar behaviors; b) set 3 covers all the target distribution for $K_0$ while the other sets are restricted to values $K_0\gtrsim 230$~MeV; c) all sets show a similar result for the symmetry energy at saturation $J_{\rm sym, 0}$ and are pushed to the lower limit of the target; d) concerning the PNM pressure sets 1 and 2 are pushed to the upper (lower) values of $P_1$ ($P_3$) while the opposite is true for set 3.

\begin{figure}[!t]
\includegraphics[width=1.\linewidth]{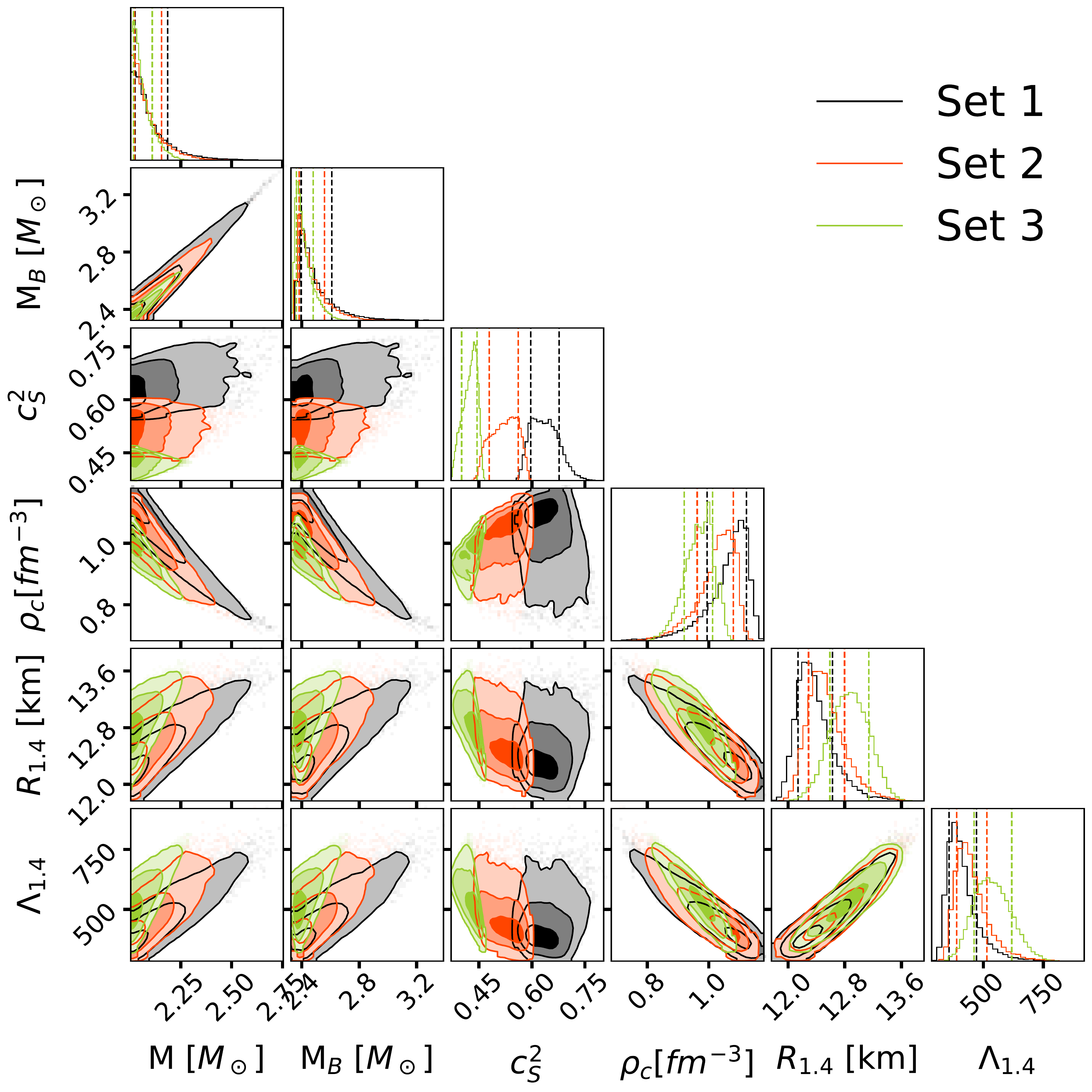}
\caption{Corner plot for the three sets of models with $\xi\in[0,0.004]$ (solid black lines), $\xi\in[0.004,0.015]$ (red) and $\xi\in[0.015,0.04]$ (green) comparing the respective NS properties, in particular, the  gravitational and baryonic maximum masses  $M_{\rm max}$ and $M_{\rm B,max}$,  the square of the speed of sound, the central baryonic density of the maximum mass configuration, and the radius and dimensionless tidal deformability of a 1.4$M_\odot$ star. The vertical lines represent the 68\% CIs, and the light and dark intensities represent the 1$\sigma$, 2$\sigma$, and 3$\sigma$ CIs, respectively.\label{T:fig5}}
\end{figure}

\begin{figure}
\includegraphics[width=0.8\linewidth]{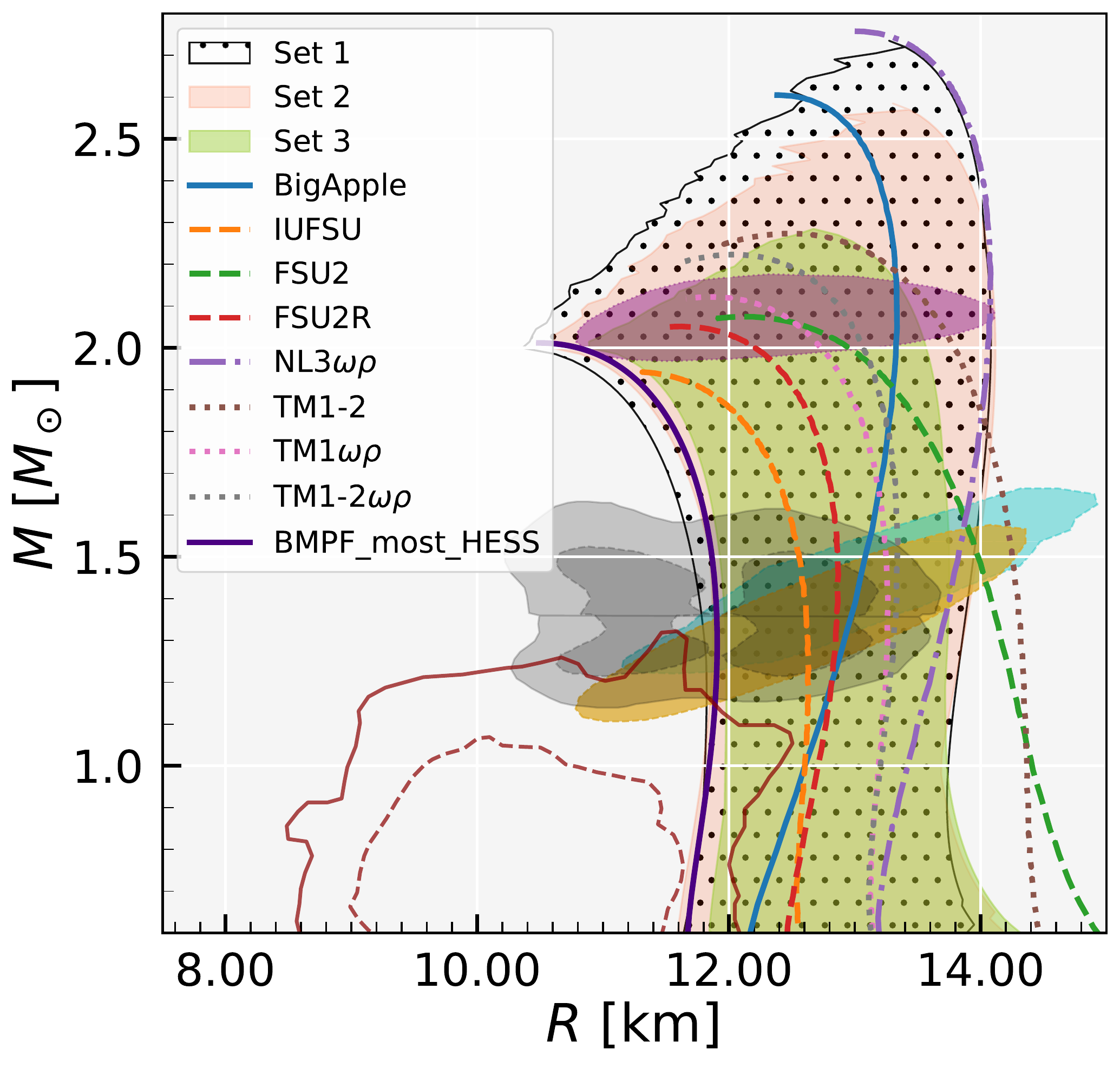}
\caption{{NS mass-radius domains 
(full posterior) produced in the following three scenarios: 
set 1 with $\xi\in[0,0.004]$ (black dot), set 2 with $\xi \in[0.004, 0.015]$ (salmon), and  set 3 with $\xi \in[0.015,0.04]$ (green).  The gray lines depict the constraints from the binary components of GW170817, along with their 90\% and 50\% credible intervals (CI).  The $1\sigma$ (68\%) CI for the 2D posterior distribution in the mass-radii domain for millisecond pulsar PSR J0030 + 0451 (cyan and yellow) 
\cite{Riley:2019yda, Miller:2019cac} as well as PSR J0740 + 6620 (violet) \cite{Riley:2021pdl, Miller:2021qha}  from the NICER x-ray data are also shown. Additionally, we show the constraint obtained from HESS J1731-347 for 68.3\% (95.4\%) CIs in dashed dark red (solid dark red) \cite{hess}.  MR curves from a few well-known RMF models are also plotted (see text for details). 
{Also, shown is BMPF$\_$most$\_$HESS, the EoS from our complete set that best describes HESS J1731-347.}
}
\label{T:fig10}}
\end{figure}

 The corner plot for the nuclear matter properties, Fig. \ref{T:fig4}, confirms the above discussion, i.e.  while sets 1 and 2 have very similar properties, set 3 differs a lot from the other two: a) concerning the symmetric nuclear matter properties, set 3 presents larger values of $Q_0$ and $Z_0$, while $K_0$ shows a Gaussian distribution centered just above 200 MeV and spreading between $\sim 100$ MeV and $\sim 300$ MeV. For the other two, the distribution of $K_0$ is squeezed above 220 MeV. It should also be noted an anti-correlation between $Z_0$ and $K_0$: the lower values of $K_0$ are compensated by larger $Z_0$; b) considering the symmetry energy properties, all sets have the same $J_{\rm sym}$ distribution, but all the other properties show differences.  Set 3 takes larger values of $L_{\rm sym,0}$ and  $K_{\rm sym,0}$, and smaller of $Q_{\rm sym,0}$ and  $Z_{\rm sym,0}$. Set 3 also shows a slight positive correlation between $L_{\rm sym,0}$ and  $K_{\rm sym,0}$. Similar behavior has been shown in \cite{Vidana2009}  for a set of quite different nuclear models. Notice, however, that this correlation is not present in  sets 1 and 2. Besides also a quite strong correlation is obtained between $L_{\rm sym,0}$ and  $Q_{\rm sym,0}$ for all three sets. Finally, it is also interesting to point out the quite broad and flat distribution  of $Z_{sym,0}$ for sets 1 and 2 while for set 3 it presents a quite peaked distribution at a low value. Lower values of  $L_{\rm sym,0}$ and  $K_{\rm sym,0}$ for sets 1 and 2 are compensated with larger values for the two higher orders, $Q_{\rm sym,0}$ and  $Z_{\rm sym,0}$.
 
 Let us now  discuss the NS properties of the three sets plotted in Fig. \ref{T:fig5}. The largest gravitational masses are obtained  with set 1. In particular, within set 1 there is a small subset for which the mass is above {2.5$M_\odot$} and as high as {2.75$M_\odot$}.  One property that distinguishes clearly the three sets is the speed of sound in the center of the maximum mass star: for set 1 the square of this quantity takes values above $~0.6c^2$, for set 3 values below $0.45 c^2$ and set 2 fill the gap between the other two distributions. 
 
 Set 3 presents the largest radius  and tidal deformability for 1.4$M_\odot$ stars and the smaller central baryonic densities indicating a stiffer EOS. Notice, however, that the small subset of models of set 1 with a mass {above} 2.5$M_\odot$ also have $R_{1.4}\gtrsim13$km and {$\Lambda_{1.4}\gtrsim 700$}. Besides, they present {a large central speed of sound, $c_s^2\sim 0.7 c^2$,} and the smallest central baryonic densities, $<0.8$fm$^{-3}$. 
 
 The baryonic and gravitational masses of the maximum mass configurations are strongly correlated. Besides, the  maximum gravitational  mass also shows a strong correlation with the radius and the tidal deformability of a 1.4$M_\odot$ NS, the larger the maximum mass the larger these two properties, and an anti-correlation with the central baryonic density of the maximum mass configuration, with larger densities associated with smaller radii and tidal deformabilities. Similar correlations have been obtained in \cite{Malik:2022jqc} and \cite{Beznogov:2022rri}, with models with density-dependent couplings.

\begin{figure*}
\includegraphics[width=0.95\linewidth]{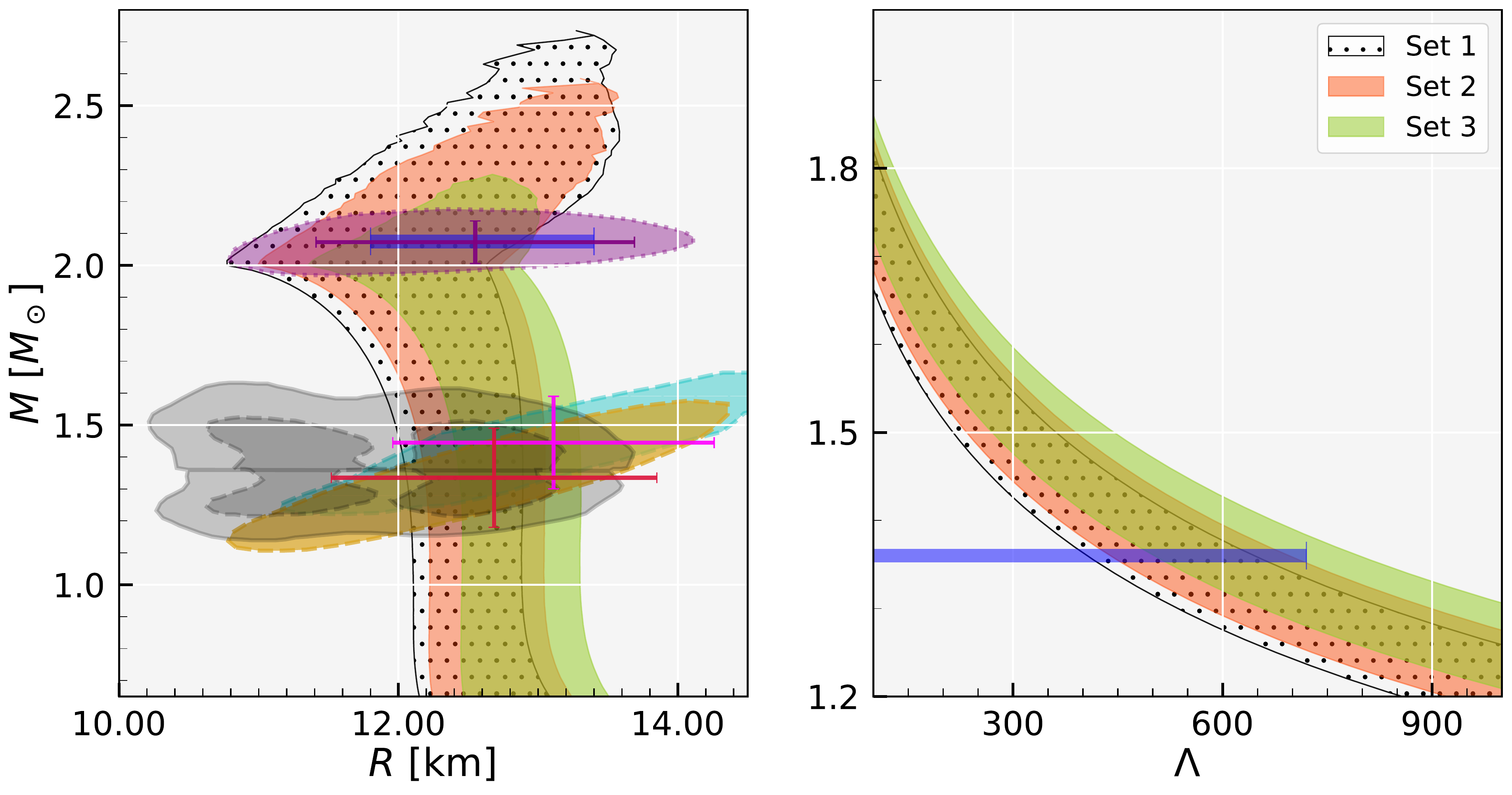}
\caption{The 90\% CI region for the sets:  $\xi\in[0,0.004]$ (black dot), $\xi \in[0.004,0.015]$ (salmon), and $\xi \in[0.015,0.04]$ (green) derived using the conditional probabilities $P(R|M)$ (left) and $P(\Lambda|M)$ (right). The gray zones in the left panel indicate the 90\% (solid) and 50\% (dashed) CI for the binary components of the GW170817 event \cite{LIGOScientific:2018hze}, for the $1\sigma$ (68\%) credible zone of the 2-D posterior distribution in mass-radii domain from millisecond pulsar PSR J0030+0451 (cyan and yellow) \cite{Riley:2019yda,Miller:2019cac} as well as PSR J0740 + 6620 (violet) \cite{Riley:2021pdl,Miller:2021qha} are shown for the NICER x-ray data. The horizontal (radius) and vertical (mass) error bars reflect the $1\sigma$ credible interval derived for the same NICER data's 1-D marginalized posterior distribution.
The blue bars depict the radius of PSR J0740+6620 at 2.08$M_\odot$ (left panel) and its tidal deformability at 1.36 $M_\odot$ (right panel) \cite{LIGOScientific:2018cki}. \label{T:fig1zeta0}}
\end{figure*}

\begin{figure}
    \centering
    \includegraphics[width=.9\linewidth]{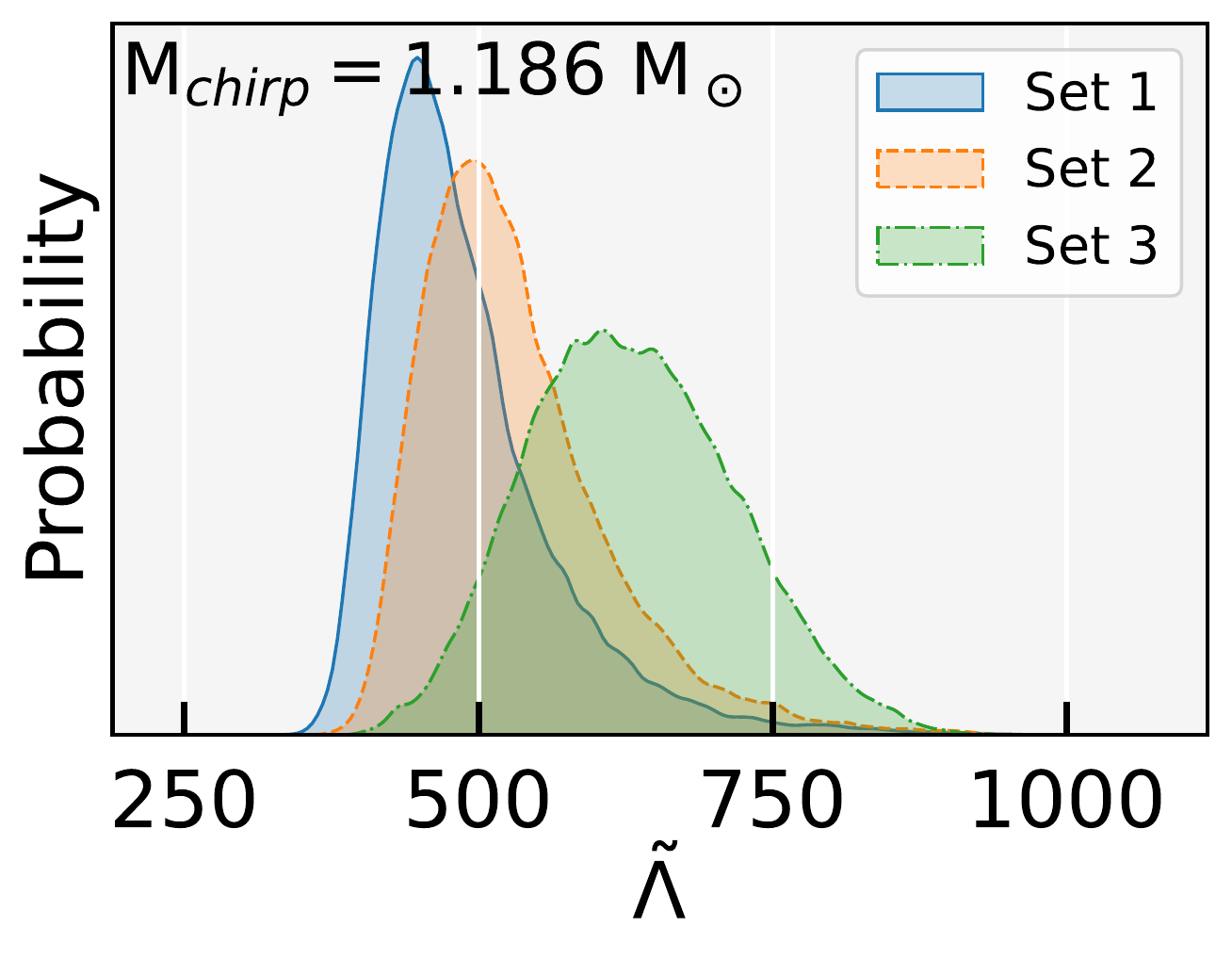}
    \caption{The Probability distribution of combined tidal deformability $\tilde \Lambda$ in a Binary is plotted for a given {\it chirp} mass M$_{chirp}=1.186~{\rm M}_\odot$ and marginalized over NS mass ratio $q=m_1/m_2$ obtained in Set 1, 2 and 3. The median and 90\% CI for $\tilde\Lambda$ are $471_{-71}^{+163}$, $516_{-84}^{+166}$, and  $626_{-132}^{+154}$ for Set 1, 2 and 3 , respectively.
    }
    \label{fig:lambr}
\end{figure}

A comparison of the NS properties predicted by the three sets becomes more evident through Fig. \ref{T:fig10} where the full posteriors for the three sets are plotted together with some astrophysical observations,  the  mass-radius prediction from the GW170817 detection \cite{LIGOScientific:2018cki} and the NICER observations of the pulsar PSR J0030 + 0451 \cite{Riley:2019yda,Miller:2019cac} and of the pulsar PSR J0740 + 6620 \cite{Riley:2021pdl,Miller:2021qha}. None of the sets is rejected by the observations. The $\omega^4$ term softens the high-density behavior of the EOS, and, therefore,  {set 3 does not} describe stars above 2.3$M_\odot$. It is interesting to discuss the properties of  set 3: a strong $\xi$ softens the EOS at high densities, therefore, in order to satisfy the 2$M_\odot$ constraint this set of models has a larger $g_\omega$ coupling, see Fig. \ref{T:fig2}, that gives rise to a stiffer EOS at low  and intermediate densities. At  high densities, the $\omega^4$ term softens the EOS and it is not possible to attain very high masses.
{
In addition, we compare the mass-radius relationships obtained from a few RMF models with our results, in particular, BigApple \cite{Fattoyev:2020cws}, IUFSU, FSU2 \cite{Chen:2014sca}, FSU2R \cite{Tolos:2017lgv}, NL3$\omega\rho$ \cite{Pais:2016xiu}, TM1-2, TM1$\omega\rho$, and TM1-2$\omega\rho$ \cite{Providencia:2013dsa}. 
It should be emphasized that the posterior we have obtained for the three sets does not completely encapsulate all models, particularly FSU2, and TM1-2. This is because those models do not satisfy all the restrictions put forth in the Bayesian setup. These two are disregarded due to the $J_{\rm sym,0}$ requirement. All the others fall inside the full posterior for the NS mass-radius domain.}

{The NS mass-radius constraint obtained from HESS J17311-347 is shown in dashed dark red (solid dark red) \cite{hess}. 
{The existence of only nucleonic composition in this  star may be questionable}  since all sets lie outside the 1$\sigma$   2-D posterior distribution in mass-radius. However, there are some EOS that falls  within the 2$\sigma$ limit. {The EoS that, considering all sets,  best matches the  HESS J1731-34 1$\sigma$ (68 \% CI) data, BMPF$\_$most$\_$HESS,  is also plotted in Fig. \ref{T:fig10}. Its model parameters together with its NMP and NS properties are given in the Supplemental Material, respectively, in Tables II and III. In the Supplemental material, we also present a few selected models for NSs with maximum mass 2.2, 2.4, 2.6, and 2.75 M$_\odot$ (the extreme one), namely  BMPF220, BMPF240, BMPF260, and BMPF275, respectively.}}

In Fig. \ref{T:fig1zeta0}, we plot the 90\% CI region of the  conditional probabilities $P(R|M)$ (left) and $P(\Lambda|M)$ (right) for the three sets. The gray zones in the left panel indicate the 90\% (solid) and 50\% (dashed) CI for the binary components of the GW170817 event \cite{LIGOScientific:2018hze}. 
The NICER x-ray data predictions  for the pulsars PSR J0030+0451 and PSR J0740 + 6620 are also included, in particular,  the $1\sigma$ (68\%) confidence zone of the 2-D posterior distribution in mass-radii domain from millisecond pulsar PSR J0030+0451 (cyan and yellow) \cite{Riley:2019yda,Miller:2019cac} as well as PSR J0740 + 6620 (violet) \cite{Riley:2019yda,Miller:2019cac}. The horizontal (radius) and vertical (mass) error bars reflect the $1\sigma$ credible interval derived for the same NICER data's 1-D marginalized posterior distribution. Finally, 
the blue bars depict the radius of PSR J0740+6620 at 2.08$M_\odot$ (left panel) and its tidal deformability at 1.36 $M_\odot$ (right panel) \cite{LIGOScientific:2018cki}.  As already indicated by the full posteriors, masses above 2.3 $M_\odot$ are only obtained within set 1 {and set 2}. Sets 1 and 2 predict $\sim0.5$~km smaller radii, as we can also confirm from Table \ref{tab:prop}. Only set 3 predicts radii above 13~km at a 90\%CI. {Notice that according to sets 1 and 2 the low mass object associated with the gravitational waves  GW190814 predicted to have a mass in the range 2.5-2.67 $M_\odot$ \cite{LIGOScientific:2020zkf} could be a neutron star. The detection of masses above 2.3$M_\odot$ puts strong constraints on $\xi$.}
Concerning the tidal deformability (right panel), set 1 and 2 prediction for $\Lambda_{1.36}$, corresponding to the $q=1$ mass ratio of the GW170817 detection, lies well inside observations, while for set 3 some models lie outside this range. 

{In order to better understand how the three sets compare regarding the tidal deformability, we plot in Fig. \ref{fig:lambr}  the effective tidal deformability $\tilde\Lambda$ probability distribution  calculated for the three sets for the chirp mass associated with the GW170817, {\cal{M}}$_{chirp}=1.186\,M_\odot$. For each and every mass-radius curve, and fixing the chirp mass at 1.186$M_\odot$, we select all possible combinations of  the mass $m_1$ and $m_2$ and calculate the combined tidal deformability. For each EOS we have 44 combinations of $m_1$ and $m_2$. None of the distributions goes below 300, consistent with the findings of several studies that show that electromagnetic counterparts of GW170817, the gamma-ray
burst GRB170817A \cite{LIGOScientific:2017zic}, and the electromagnetic
transient AT2017gfo \cite{LIGOScientific:2017ync} set a lower limit on the  $\tilde\Lambda$ of the order of 210  
\cite{Bauswein:2019skm}, 300 \cite{Coughlin:2018miv}, 279 \cite{Wang:2018nye}, and 309 \cite{Radice:2018ozg}. {The median along with its 90\% CI of the three distributions corresponding to sets 1, 2, and 3 are, respectively,  $471_{-71}^{+163}$, $516_{-84}^{+166}$, and  $626_{-132}^{+154}$}. Set 3 has a quite symmetric and wide distribution while the other two are narrower asymmetric distributions that spread above the 720 limits obtained from \cite{LIGOScientific:2018cki}.  
}
\begin{figure}
\includegraphics[width=0.95\linewidth]{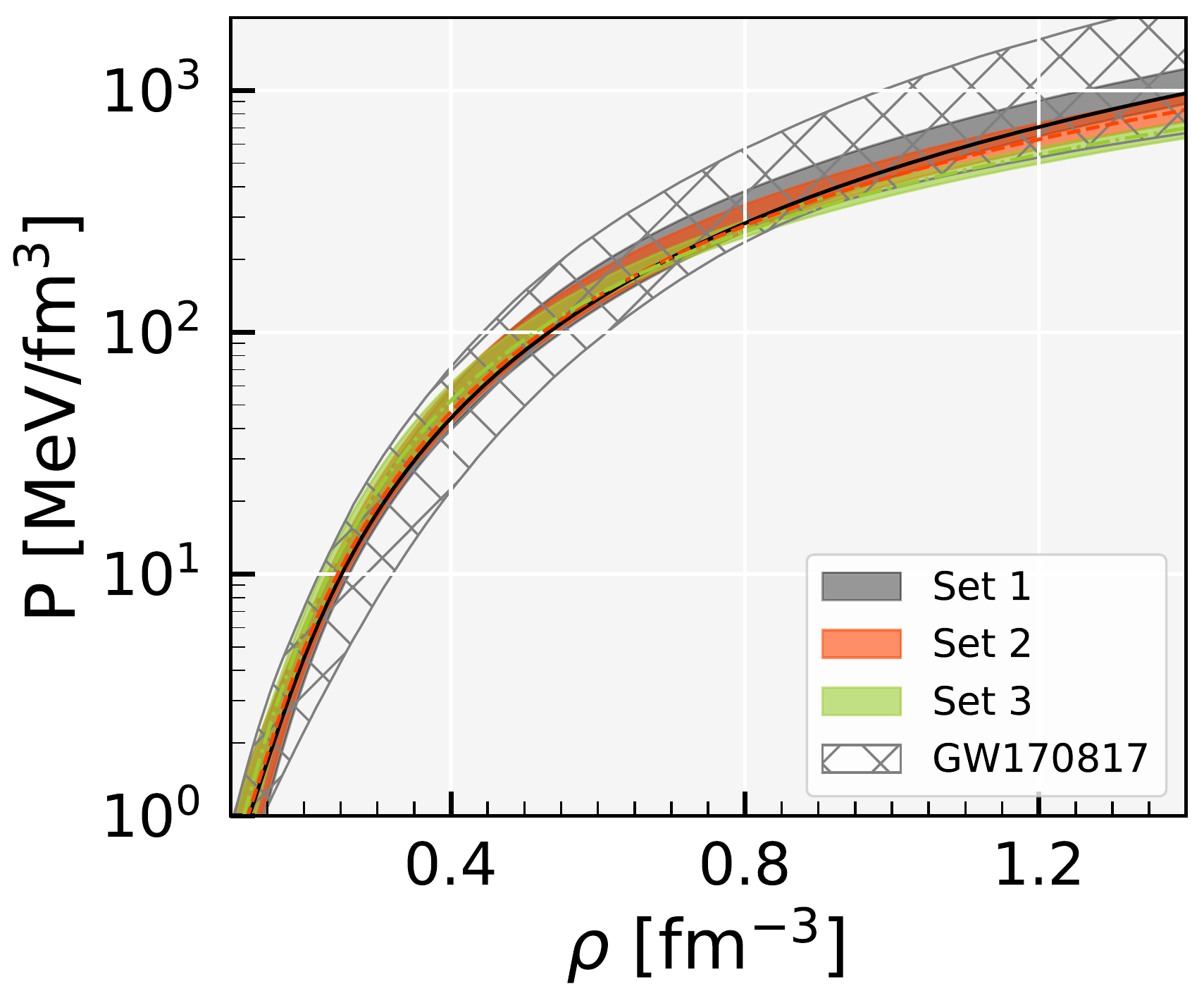}
\caption{Pressure versus the baryonic number density for the three scenarios $\xi \in[0,0.004]$ (dark grey), $\xi\in[0.004,0.015]$ (salmon) and $\xi\in[0.015,0.04]$ (green). Also shown is the band predicted from  the GW170817  event (hatched grey). \label{T:fig6}}
\end{figure}

In Fig. \ref{T:fig6}, we plot the $\beta$-equilibrium pressure as a function of the baryonic density for the three sets  ($\xi<0.004$, {$0.004<\xi <0.015$ and $\xi>0.015$}), together with the prevision obtained from GW170817  \cite{LIGOScientific:2017zic}.  All models fall inside the GW170817 band. However, their behavior can be distinguished: a smaller $\xi$ implies a softer EOS at lower densities, harder at high densities, and the other way around.

\begin{figure}[ht]
\includegraphics[width=0.95\linewidth]{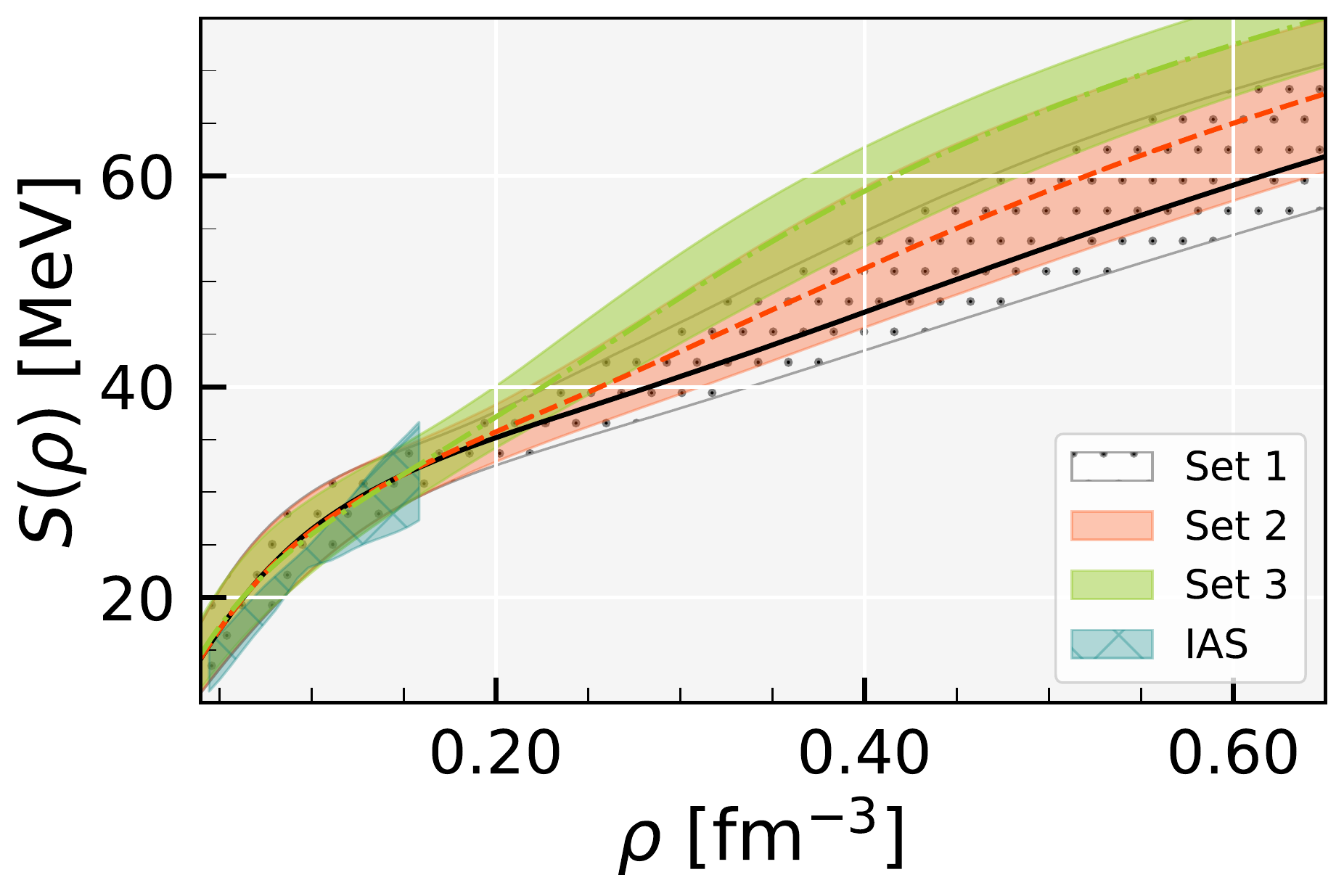}
\caption{The symmetry energy versus the baryonic number density for the three sets with $\xi\in[0,0.004]$ (dark grey), $\xi \in[0.004,0.015]$ (salmon), and $\xi \in[0.015,0.04]$ are plotted (green). The constraint depicted from the IAS analysis is also illustrated by the light sky region.\label{T:fig7}}
\end{figure}
\begin{figure}[ht]
\includegraphics[width=0.95\linewidth]{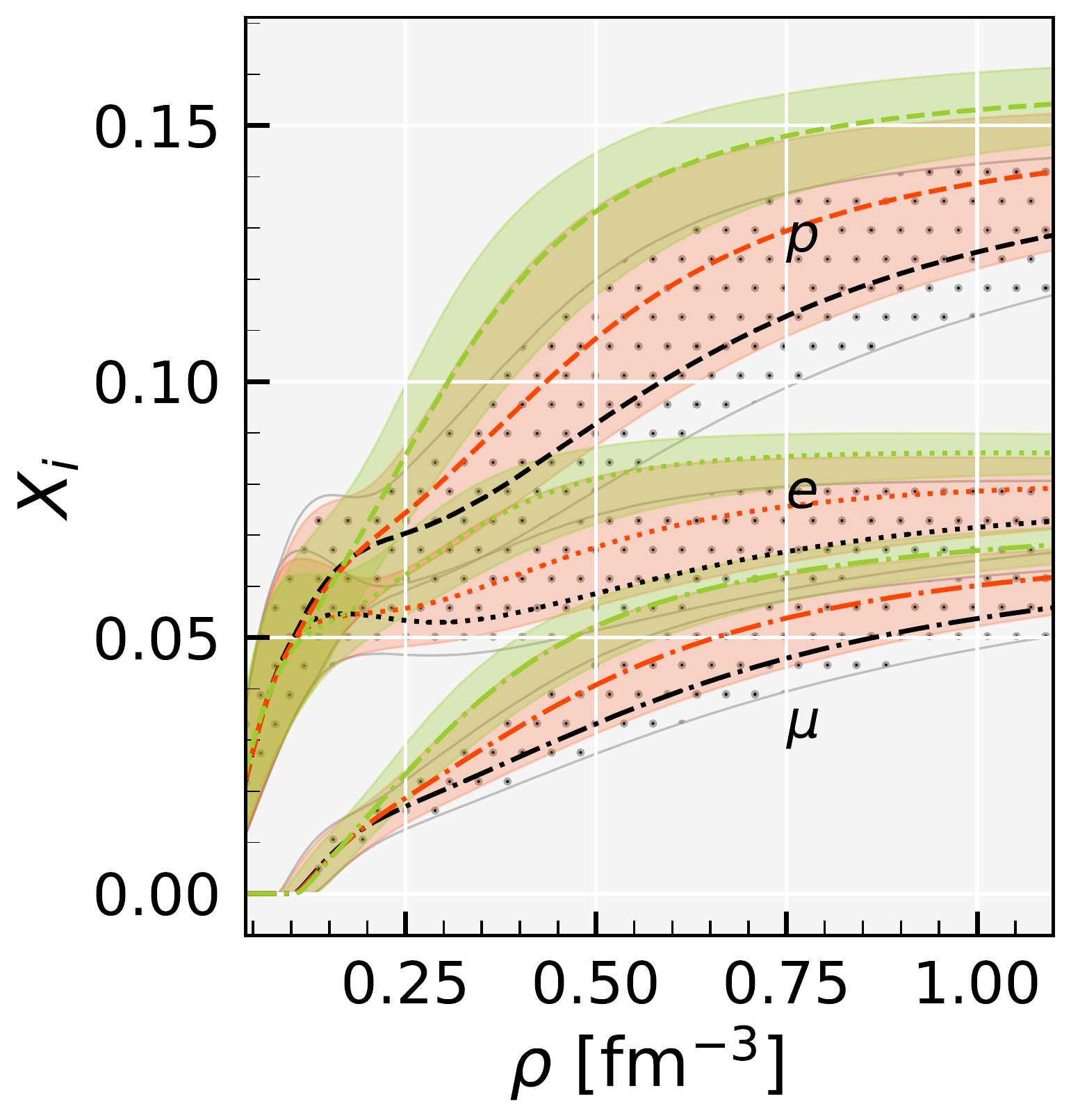}
\caption{ A comparison of the proton, electron, and muon fractions versus the baryonic density in the three different scenarios: $\xi \in[0, 0.004]$ (dark grey), $\xi \in[0.004, 0.015]$ (salmon), and $\xi \in[0.015,0.04]$ (green).
\label{T:fig8}}
\end{figure}

In Fig. \ref{T:fig7}, the symmetry energy is represented for the three scenarios considered in our study. We conclude that the larger $\xi$ the stiffer is the symmetry energy, favoring larger proton fractions as seen in Fig. \ref{T:fig8}. As referred in Sec. \ref{formalism}, a nonzero $\xi$ gives rise to a larger $\varrho$ effective mass, Eq. (\ref{mr}), therefore, having a direct influence on the strength of the $\varrho$ field. The $\omega$-field is proportional to the baryonic number density $\rho$ if \{$\xi=0$\}, while for a nonzero $\xi$, $\omega$ increases with a smaller power of $\rho$. So the larger the value of $\xi$ the smaller the $\varrho$ effective mass and the larger the $\varrho$ field.
A large $\varrho$-field gives rise to a smaller  isospin  asymmetry, i.e. larger proton fractions will occur. However, since larger proton fractions favor the direct Urca (DUrca) process inside NSs with smaller masses, the different scenarios represented by the three sets may be distinguished by their cooling properties.

\begin{figure}
\includegraphics[width=0.95\linewidth]{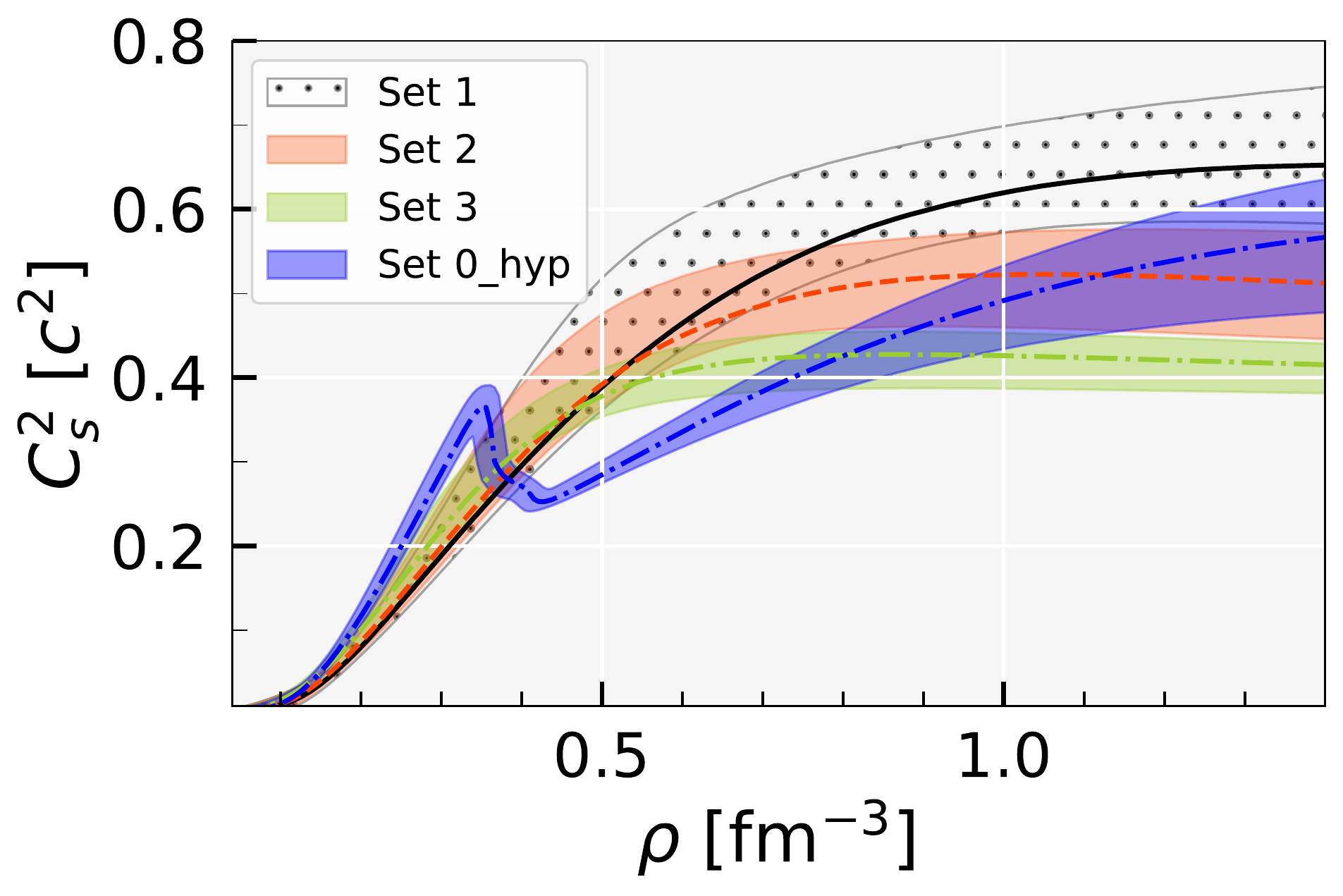}
\caption{The median and 90\% credible interval of the square of sound velocity ($c_s^2$) as a function of baryon density are shown for $\xi\in[0,0.004]$ (black dot), $\xi \in[0.004,0.015]$ (salmon), and $\xi \in[0.015,0.04]$ (green).
{The blue region represents the 90\% credible interval of the square of sound velocity allowing for the onset of hyperons in Set 0.}
\label{T:fig9}}
\end{figure}

Also very interesting is the analysis of the speed of sound behavior for the three sets. While for the $\xi<0.004$ set, the speed of sound increases monotonically with the baryonic density, this is not so for the $\xi>0.015$ set, see Fig. \ref{T:fig9}: in this case, the speed of sound square attains a maximum below 0.45$c^2$ at $\rho\sim 4\rho_0$  and then decreases smoothly. The average behavior of the set with $0.004<\xi<0.015$ shows an intermediate behavior as expected. In this last case for the densities plotted in Fig. \ref{T:fig9}, the speed of sound has stabilized just above {0.5$c^2$}. {The blue region in the figure represents the 90\% credible interval of the square of sound velocity that allows for the onset of hyperons in Set 0, as discussed below under section \ref{hypqcd}.}

\begin{figure*}
\includegraphics[width=1.\linewidth]{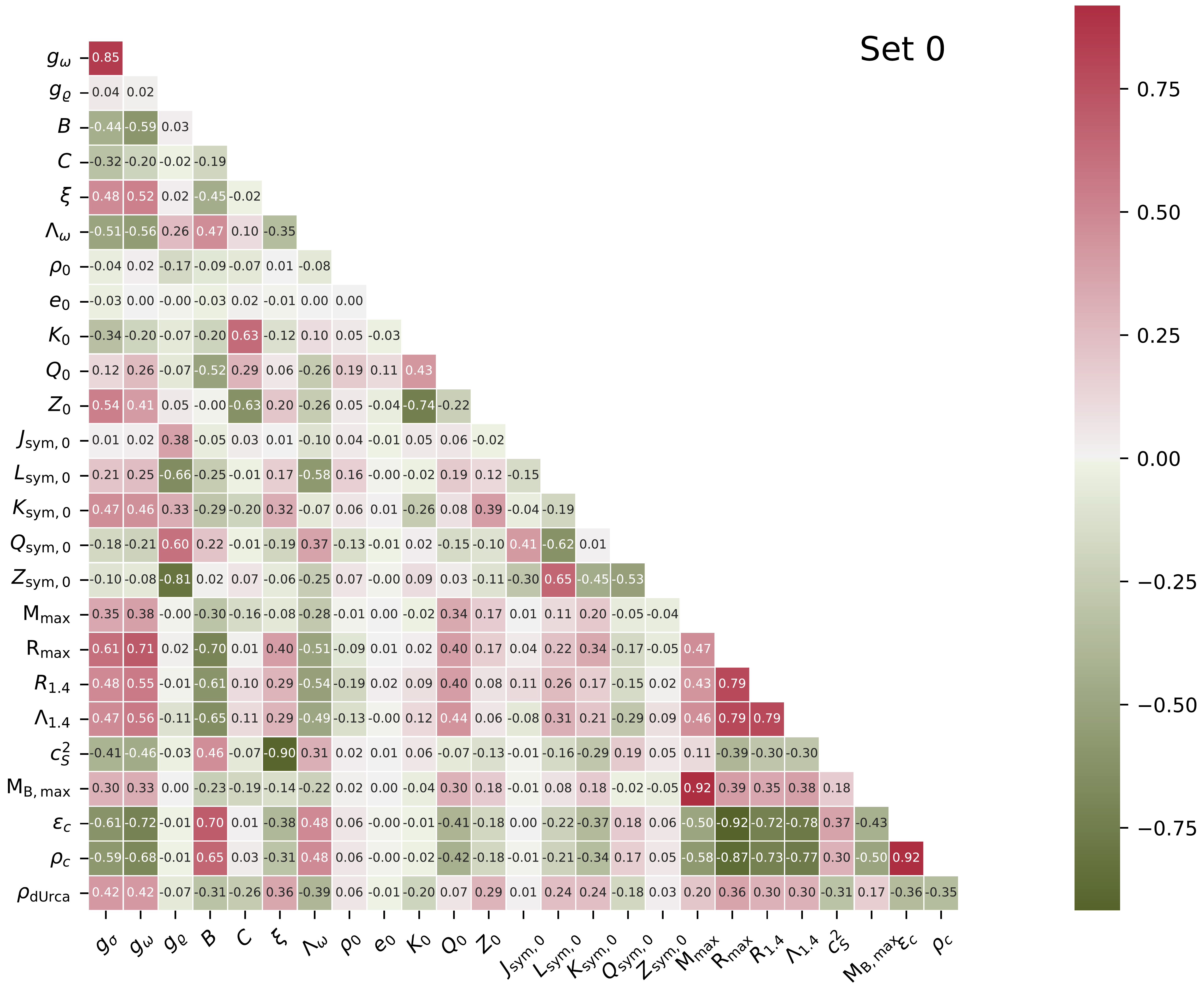}
\caption{The Kendall rank correlation coefficients between RMF model parameters, nuclear saturation properties (NMP), and neutron star properties (NS) were obtained from the posterior with prior Set 0. In such figures, Pearson's correlation coefficient is typically employed. Pearson's correlation coefficient measures a linear relationship between two variables, whereas Kendall's correlation coefficient measures a monotonic relationship. \label{cor0}}
\end{figure*}

{Finally, we study the correlations between the different quantities considered, in particular, model parameters, nuclear matter properties, and neutron stars properties, see Fig. \ref{cor0} where the Kendall rank correlation coefficients are shown for set 0.  
The strongest correlations obtained with coefficients of the order of 85\% or above are between: a) $g_\sigma$ and $g_\omega$ for which 85\% was determined. The correct description of the binding energy strongly constrain these two parameters; b) the central baryonic density and energy density of the maximum mass star with the corresponding star radius, respectively, -87\% and -92\%. This correlation was referred in \cite{Jiang:2022tps} and will be discussed below; c) the speed of sound in the center of the maximum mass star with the parameter $\xi$, -90\%.   This correlation reflects the fact that the parameter $\xi$ determines the stiffness of the EoS at high densities; d) the gravitational mass of the maximum mass star with the corresponding baryonic mass, 92\%, and the central baryonic density  with the energy density of the maximum mass star, also 92\%. 
}

{ As discussed above, the correlation coefficient between the central density of the maximum mass star $\rho_c$ and its radius $R_{\rm max}$ is of the order of $0.9$, see Fig. \ref{cor0}. A similar result was obtained in \cite{Jiang:2022tps} with a set of  EoS determined using the sound-speed parameterization method and constrained  to satisfy  X-ray and gravitational-wave observations, and ab-initio calculations, in particular, low-density neutron matter chiral effective theory  and  high density perturbative QCD results.  These authors found that the normalized central density of the maximum mass star was related to the  corresponding radius through the quadratic relation
 $$
\frac{\rho_{c}}{0.16 {~~\rm fm}^{-3}}=d_0\left[1-\left(\frac{R_{\rm max}}{10 \mathrm{~km}}\right)\right]+d_1\left(\frac{R_{\rm max}}{10 \mathrm{~km}}\right)^2,
$$
with $d_0 = 27.6$ and $d_1 = 7.5$ and a 3.7\%  standard deviation {of relative residual} over the central value zero. Performing a similar analysis with Set 0, we have obtained
$d_0=28.89 \pm 0.02$ and $d_1=7.73 \pm 0.01$.}
The parameters  $d_0$ and $d_1$ obtained with our approach  and in \cite{Jiang:2022tps} differ less than 5\% although very different EOS descriptions have been used. 
{Notice, however, that the linear relation shows a  chi-square fit similar to the quadratic relation. We have obtained  for Set 0
$$
\frac{\rho_{c}}{0.16 {~~\rm fm}^{-3}}=m_0 \left(\frac{R_{\rm max}}{10 \mathrm{~km}}\right)+c_0,
$$
with $m_0=-11.618 \pm 0.018$ and $c_0=19.255 \pm 0.019$.
The relative residual for $\rho_c$ with Set 0 data, obtained with  both non-linear and linear relations shows a symmetric Gaussian distribution centered over zero with 1.4\% standard deviation. 
}

\section{The Hyperons and perturbative QCD \label{hypqcd}}
{In this section we complete the discussion of the previous section by addressing two issues frequently considered: a) how will non-nucleonic degrees of freedom affect the conclusions; b) are the constraints obtained from pQCD for densities as the ones found inside neutron stars affecting the present neutron star  description? The two topics will be discussed in the following subsections.}

\subsection{Effect of hyperons\label{hyperons}}
\begin{figure}
\includegraphics[width=1.\linewidth]{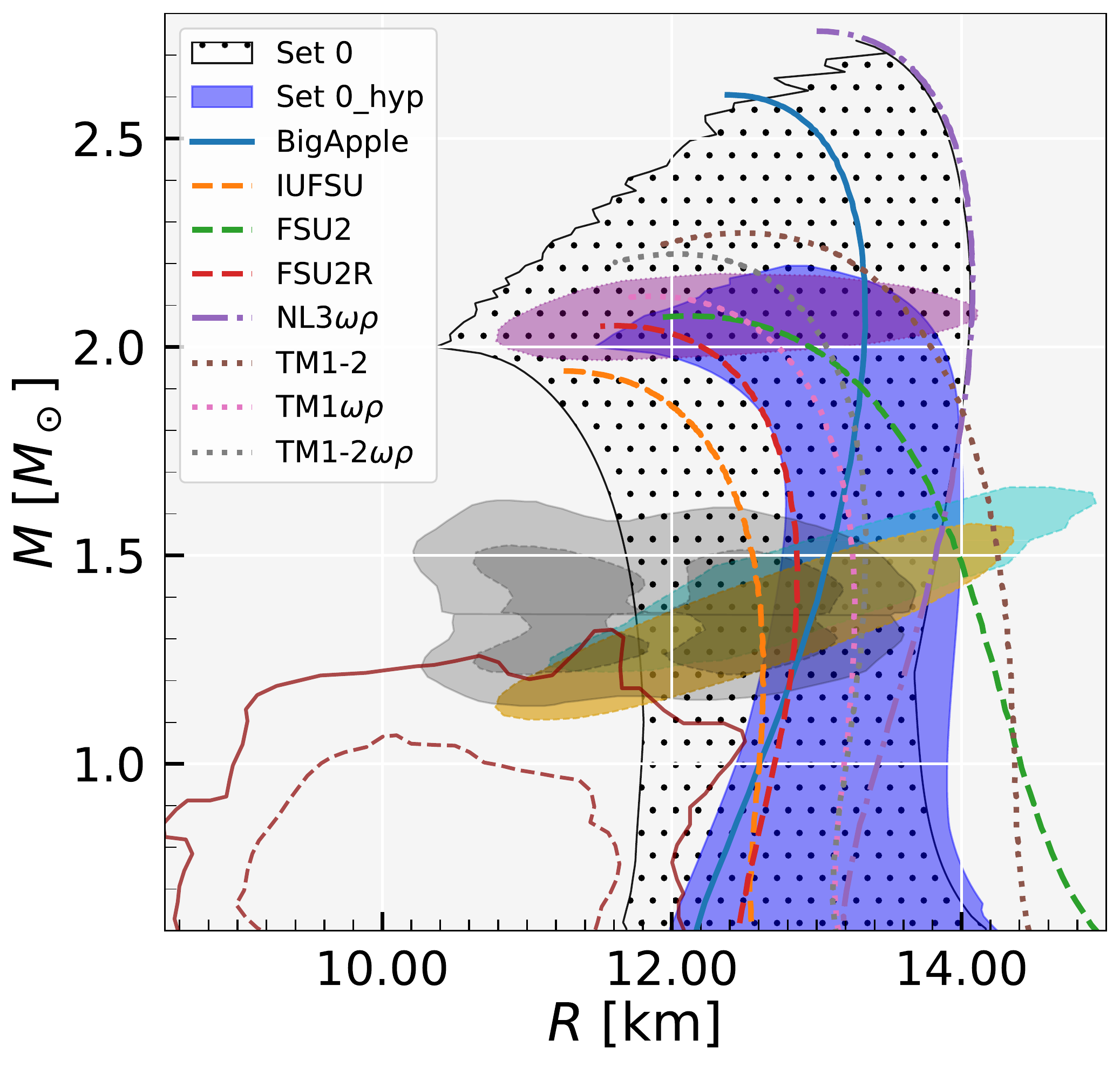}
\caption{{The NS mass-radius complete domains generated using only nucleonic and hyperonic matter, based on the Set 0 conditions, are depicted in their full posterior form: the dotted (blue) region corresponds to the no-hyperon (hyperon) calculation. For the meaning of all the other regions and curves please see  the caption of Fig. \ref{T:fig10}} 
\label{mrhyp}}
\end{figure}

{The appearance of hyperons in neutron stars, or other nucleonic degrees of freedom,  is an open question in astrophysics and is still the subject of ongoing research.  For instance, in \cite{Lonardoni:2014bwa} the authors conclude  within an auxiliary field diffusion Monte Carlo description of nuclear matter with $\Lambda$-hyperons the onset of hyperons is very sensitive to the three-body force, and may disfavor the onset of hyperons. However, if hyperons are considered in a RMF description of neutron star matter, the onset of hyperons generally occurs for densities of the order $2-3 \rho_0$.}

{We will introduce hyperons following the approach described in \cite{Malik:2022jqc}. The interaction between nucleons and hyperons is defined by the $\sigma$, $\omega$, $\rho$, and $\phi$ mesons, and we allow for the possible onset of the  neutral $\Lambda$-hyperon  and the negatively charged $\Xi^-$-hyperon.  The $\Lambda$-hyperon generally sets in first  and  the $\Xi^-$-hyperon secondly \cite{Weissenborn:2011kb,Fortin:2016hny,Fortin:2020qin}. The $\Sigma$-hyperon potential in the nuclear matter is possibly repulsive  disfavoring the onset of this hyperon before the $\Xi^-$-hyperon, see \cite{Gal2016}. We consider that the coupling of the hyperons to the vector-isoscalar mesons ($\omega$ and $\phi$-mesons) is determined by the SU(6) symmetry
\bea
	&&g_{\omega\Lambda}=\frac{2}{3} g_{\omega N}  \ , \quad  g_{\omega\Xi^-}=\frac{1}{3} g_{\omega N}  \\ 
&& g_{\phi\Lambda}= - \frac{\sqrt{2}}{3} g_{\omega N}, \,  \quad g_{\phi\Xi^-} - \frac{2\sqrt{2}}{3} g_{\omega N}
		\label{eq:SU6-relation2}
\eea
and for the $\rho$-meson we assume 
\bea	
g_{\rho B} = g_{\rho N}  
	\label{eq:SU6-relation3}
\eea
In the Lagrangian density, the interaction term between the $\rho$-meson and baryons takes into account the isospin explicitly.
The coupling of the $\sigma$-meson to the baryons is  written in terms of the coupling to the nucleon as $g_{\sigma Y}=x_{\sigma Y}\, g_{\sigma N}$, with $x_{\sigma Y}$ fitted to hypernuclei properties. Considering several models,  the factor $x_{\sigma\Lambda}$ takes values between 0.609 and 0.622, and  values between 0.309 and 0.321 were calculated for $x_{\sigma\Xi^-}$. These two intervals have been used in the calculation with hyperons. 
The same prior  used to define set 0 (see Table \ref{tab2})  together with the above intervals for the baryon-$\sigma$ meson were considered, as well as  the constraints defined in Table \ref{tab1}.}

 {The effect of the inclusion of hyperons on the total mass-radius domain span by the hyperon EoS set is plotted in Fig. \ref{mrhyp}. The maximum mass that is attained has reduced from 2.7 $M_\odot$ for nucleonic stars to $\sim 2.2 M_\odot$ for hyperonic stars.  A strong effect is also observed on the radius: the smaller radius region was eliminated, and, simultaneously the mass-radius region extends  to slightly larger radii. The  EoS has to be stiffer in order to be able to describe 2$M_\odot$ stars. The EoS obtained are characterized by a very small value of $\xi$ (the median is  0.00137,  and the 68\% CI is [0.0004,0.00326]) as expected because a large $\xi$ softens the EoS at high density disfavoring the possible description of stars with a mass equal or above 2$M_\odot$.}

{The behavior of the speed of sound in the presence of hyperons is shown in Fig. \ref{T:fig9}  where it can be compared with the no-hyperon calculation.  The hyperon onset has a strong effect on the speed of sound as discussed in other works \cite{Malik:2022jqc}.  The speed of sound presents a maximum at the onset of hyperons, for a density close to the one predicted in \cite{Gorda:2022jvk} and \cite{Altiparmak:2022bke} with an agnostic description of the EoS. Agnostic descriptions, however,   do not allow the determination of the star composition \cite{Annala2019,Annala:2021gom,Altiparmak:2022bke,Somasundaram:2022ztm}. } 

{We conclude that the description of neutron star matter based on the microscopic model of nuclear matter considered shows a behavior of the speed of sound compatible with the results of \cite{Gorda:2022jvk} and \cite{Altiparmak:2022bke}: in our framework, if the parameter $\xi$ is large enough the speed of sound increases until a value of $\sim 0.4c-0.45c$ at $\sim 3\rho_0$ and then  stabilizes or decreases.   If in the future the speed of sound is constrained and a speed of sound of the order of $0.4 c$ is obtained in the center of a NS, the present work shows that we do not necessarily need exotic degrees of freedom or a deconfinement phase transition to interpret this value.}
{Note, however, that  although our results are compatible with the prediction of \cite{Gorda:2022jvk}, where the authors only  give a 68\% CI,  there are some qualitative differences, in particular, concerning the sharp peak of the sound speed  followed by a softening that occurs at 4$n_s$. The sharp peak is missing in our analysis with nucleonic matter, and it exists but with a  different structure in our study with hyperonic matter. Besides, the low density constraints imposed in both works are different. Since no confidence interval is given in \cite{Hebeler2013}, in our work we have taken the uncertainty to be the double of the one given motivated by the dispersion of chEFT results compiled in \cite{Raaijmakers:2021uju}, and this was not considered in \cite{Gorda:2022jvk}: the larger uncertainty will result in a wider band at low densities.}

\begin{figure*}
\includegraphics[width=1.\linewidth]{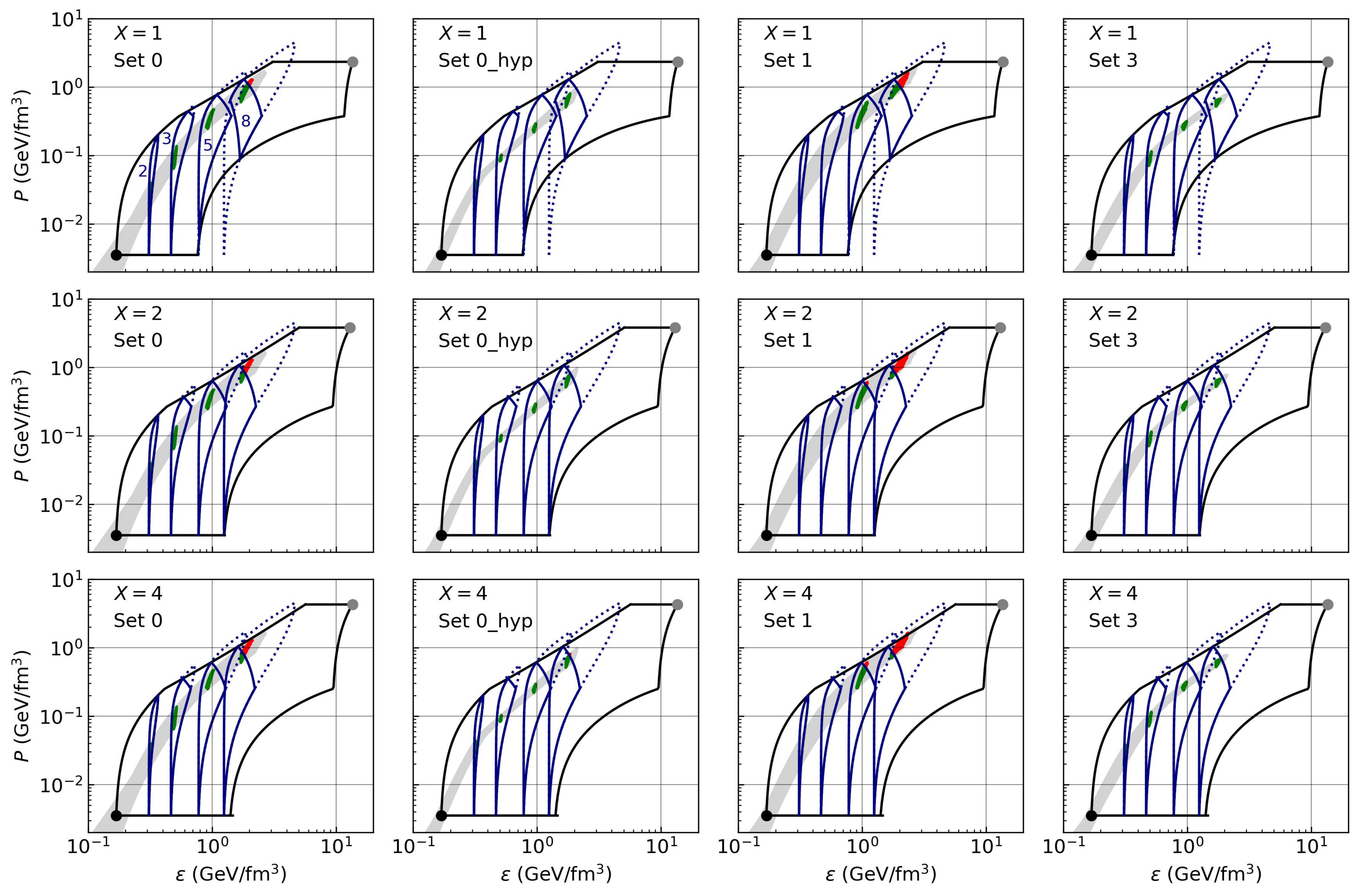}
\caption{{We display the values of energy density ($\varepsilon$) and pressure (p) for four different sets: Set 0, Set 0 with hyperon,
Set 1, and Set 3 (from left to right columns). In addition, we apply the robust equation of state constraints from Ref. \cite{Komoltsev:2021jzg} that ensures stability, causality, and thermodynamic consistency. The regions enclosed by the solid blue lines are subject to pQCD constraints that restrict the values of energy density ($\varepsilon$) and pressure (p) within the same solid line regions. These constraints apply specifically to baryon number densities of $n = 2,$ 3, 5, and 8 $n_s$ (where $n_s$=0.16 fm$^{-3}$). In contrast, dotted blue lines represent excluded regions,
{where not all pQCD conditions are met.}
We show the results for different renormalization scale parameters X \cite{Kurkela2009} for 1, 2, and 4 in order from top to bottom row. The green and red dots represent, respectively, the models in our sets that satisfy and do not satisfy pQCD constraints.}}
\label{pQCD}
\end{figure*}

\subsection{Effect of pQCD constraints}
{
In our Bayesian inference, no constraints from pQCD have been included. Our framework is not valid for densities as high as the ones explored with pQCD,  however,  indirect constraints may be imposed. It has been shown in \cite{Komoltsev:2021jzg,Gorda:2022jvk} that pQCD constraints have a finite effect at densities found inside NS, and in \cite{Komoltsev:2021jzg} a set of constraints on the pressure, chemical potential and baryonic density were calculated using information from thermodynamic potentials together with causality and stability conditions.  In this subsection, we discuss the compatibility of our different EoS sets with the pQCD constraints deduced in \cite{Komoltsev:2021jzg}.}

{In Fig. \ref{pQCD} we plot for three values of the QCD scale X the pressure versus energy density including the pQCD constraints on these quantities, respectively from left to right, for set 0,  set 0 with hyperons, set 1 and set 3. We also identify the constrained region for selected baryonic densities, up to 8 times saturation density taking for this quantity a reference value, $n_s=0.16$~fm$^{-3}$. This density is above the central density  of the maximum mass star of all our sets.  Set 1 is having the largest densities in the center and at 90\% CI these are below 7.2$n_s$.
At 5$n_s$ almost all models satisfy pQCD. However, at $8n_s$ some models fail the constraints, in particular, some models with a small $\xi$, depending also on the  value of QCD scale X.  It is interesting that all models of set 3 (large values of $\xi$) satisfy the pQCD constraints independently of the scale. Also, the set that includes hyperons essentially satisfies the pQCD constraints. In the future, these constraints could be imposed in the Bayesian inference. 
{As can be seen,} at high densities, $X=4$ is imposing the strongest constraints. Models of Set 1 (with the smaller $\xi$) are the ones that fail more frequently the constraints at 8~$n_s$. Considering the constraint $X=1$ in set 1, from the total 21037 models 618 do not satisfy the pQCD constraints. The last models have larger maximum masses ($2.307 - 2.512\,M_\odot$ at 90\% CI in contrast with $2.07 - 2.51 M_\odot$ for the models that satisfy the constraints). The absolute maximum mass is $\sim 2.75 M_\odot$ for the excluded models and $\sim 2.5M_\odot$ for the others.  }

\begin{figure}
\includegraphics[width=0.95\linewidth]{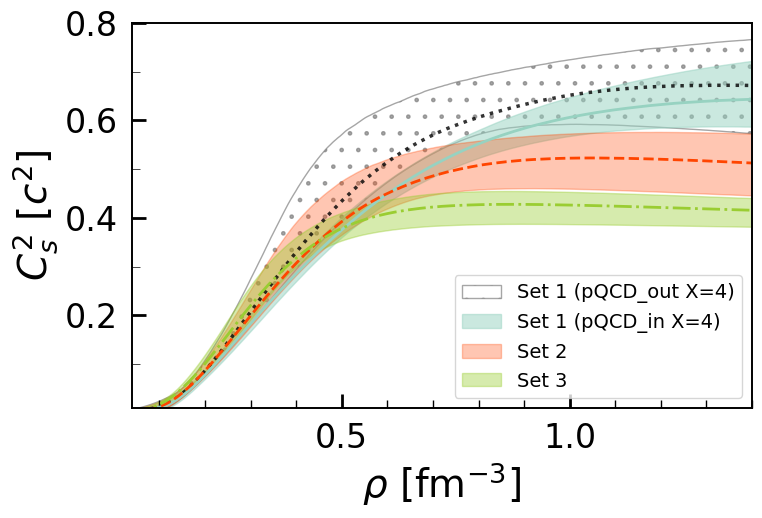}
\caption{The effect of pQCD constraints on the speed of sound: the median and 90\% credible interval of the square of the speed of sound ($c_s^2$) as a function of baryon density are shown for Set 1 ($\xi\in[0,0.004]$) pQCD excluded EOS (pQCD\_out, black dot) and pQCD included EOS (pQCD\_in, light blue), for Set 2 ($\xi \in[0.004,0.015]$, salmon), and for Set 3 ($\xi \in[0.015,0.04]$, green).
\label{T:fig12}}
\end{figure}
{In order, to understand which models do not satisfy pQCD constraints we have considered the combination that excludes the largest number of EOS, Set 1 with the QCD scale $X=4$, to allow for acceptable statistics. Under these conditions 12662 models satisfy  the pQCD  constraints (pQCD\_in) and 8375 do not (pQCD\_out). In Fig. \ref{T:fig12}, the speed of sound of the two sets pQCD\_in and pQCD\_out of Set 1 is compared with the other two sets, Set 2 and Set 3,  shown in Fig. \ref{T:fig9}. Models excluded by pQCD have in average  the largest speeds of sound at all densities.
A large number of the excluded models have a $\xi$ parameter  close to zero, but not all of them, and on average a larger $g_\omega$ coupling and a larger incompressibility. As a result, larger radii are predicted for $1.4 M_\odot$  stars, as well as larger maximum masses,  and smaller baryonic densities at the center of the maximum mass star, corresponding to harder EOS.}

\section{Conclusions \label{con}}
In the present study, we have studied the nuclear matter properties and NS properties obtained within a RMF description of nuclear matter. We have considered a RMF model that includes mesonic non-linear self-interaction and mixed interaction terms as in models discussed in \cite{Boguta1977,Sugahara:1993wz,Horowitz:2000xj,Todd-Rutel2005,Chen:2014sca}. A Bayesian inference analysis was performed, considering flat distributions for the priors, the model parameters, and imposing a small number of nuclear matter properties and the 2$M_\odot$ observational constraint.  

Presently, nuclear matter properties at saturation are reasonably well constrained, however, at high densities there is still too little information to constrain nuclear models. In the RMF model used in our study, the non-linear $\omega^4$ term has a special role in establishing the high-density behavior of the EOS.  We have, therefore, considered three different scenarios by imposing different constraints to the coupling  $\xi$ of the $\omega^4$ term.

One of the main conclusions is that the strength of the $\omega^4$ term controls the magnitude of the speed of sound in the center of the star: a larger coupling will originate a smaller speed of sound in the center. However, a smaller speed of sound also indicates a softer EOS at high densities. The two solar mass constraint in this model with a large coupling $\xi$ is only satisfied if the EOS is stiff at low and intermediate densities, and, therefore gives rise to larger 1.4$M_\odot$ NS radii.   At 90\% CI we have obtained for the 1.4$M_\odot$ star the radius $11.99< R_{1.4}<12.66$ km for $\xi<0.015$ which increases to $12.44< R_{1.4}< 13.29$ km if $\xi>0.015$ is considered.

It is interesting to verify that for set 3 ($\xi>0.015$) the speed of sound  has a non-monotonous behavior: it attains a maximum around $4\,\rho_0$ and decreases for larger densities. In \cite{Gorda:2022jvk,Kurkela:2022elj}, the authors study the behavior of the speed of sound at high density extrapolating the equation of state to high densities using a Gaussian process EoS description. They  condition the EOS  to astrophysical observations, or to both astrophysical observations and pQCD, and verify that the QCD conditioning gives rise to a decrease of the speed of sound above $\sim3 ~\rho_0$ after a steep rise until this density.  Notice that this is precisely the density at which the speed of sound of the three sets cross in Fig. \ref{T:fig9}. The softer the EOS above that density, the stiffer  it is  below this reference density and the other way around.  {The decrease of the speed of sound with the onset hyperons, as discussed in Sec. \ref{hyperons} and in \cite{Malik:2022zol},  occurs below 3$\rho_0$ but for values of $c_s^2$ of the same order of magnitude $\lesssim 0.4 c^2$.  The probability distribution for sets 2 and 3 and set 0 with hyperons in Fig. \ref{T:fig9} are compatible with the results of \cite{Gorda:2022jvk} when pQCD constraints are imposed.  If in the future the speed of sound is constrained and a speed of sound of the order of 0.4 $c$ is obtained in the center of a NS, the present study shows that it is not necessary to include exotic degrees of freedom or a deconfinement phase transition to interpret this value.
}

All observational constraints existing presently (from NICER, from LIGO-Virgo Collaboration, and from the measurement of NS masses above two solar masses) can be satisfied within the RMF model discussed.  Notice that the GW170917 tidal deformability constraint is well satisfied by the present model. The maximum mass attained is {$\sim 2.75\,M_\odot$} and was obtained for a $\xi<0.004$, i.e. for an almost zero $\omega^4$ term. For a finite $\xi>0.015$ the maximum mass obtained is $\sim 2.3M_\odot$.

Another important nuclear matter property affected indirectly  by the $\omega^4$  term is the symmetry energy. It was discussed that a larger  $\omega^4$ term gives rise to a larger $\varrho$-field and, therefore, a smaller proton-neutron asymmetry. A direct effect is the onset of direct Urca nucleonic processes at lower densities, and, therefore smaller NS masses.

{We have also confirmed the anti-correlation obtained in \cite{Jiang:2022tps} between the maximum mass radius and the corresponding central baryonic density with a set of EOS built using the speed of sound method. We have shown that both a linear and a quadratic relation give rise to  a similar  chi-square fit.}

It is interesting to establish a comparison with the results of a similar  Bayesian inference analysis carried out in a different family of RMF models in \cite{Malik:2022jqc}, where a model with density-dependent couplings was considered. The  high-density behavior of the EOS in our approach is defined by the non-linear meson terms included in the Lagrangian density, which are not included in the formulation with density-dependent couplings.  Comparing the outputs in both studies we conclude that the conclusions drawn in \cite{Malik:2022jqc} do not differ  much from the results obtained with set 2. Set 1 predicts larger maximum masses and speed of sound than the ones obtained in \cite{Malik:2022jqc}. On the other hand, set 3 predicts larger radii for the canonical NS and smaller central speeds of sound, clearly showing a different high-density behavior. 

In \cite{Malik:2022zol}, the authors undertook the Bayesian inference considering  the possibility that hyperons nucleate inside NS. In that study, the authors concluded that the joint effect of the presence of hyperons and the two solar mass constraints was the prediction of larger radii for intermediate mass NS. This is a conclusion similar to the one drawn with set 3: the $\omega^4$ softens the EOS, in an equivalent way the onset of hyperon does, and, as a consequence the EOS has to be stiffer at intermediate densities, giving rise to larger radii. {We have also studied the onset of hyperons in the present framework. The two solar mass constraint restricts the parameter $\xi$ to quite small values. On average the NS radius of a 1.4$M_\odot$ star increases and the speed of sound has a steep drop around 2$\rho_0$ and a moderate growth for larger baryonic densities keeping inside the range constrained by pQCD \cite{Gorda:2022jvk}. }

{It has been shown in \cite{Komoltsev:2021jzg} that pQCD imposes constraints at densities that can be as low as $\sim 2 n_s$. We have verified whether the different EoS sets generated satisfy the constraints deduced in  \cite{Komoltsev:2021jzg} and concluded: a) the constraints are satisfied for any QCD scale $X\in[1,4]$ if a $\xi>0.015$ is used; b) the set with hyperons and any value of $\xi$ satisfies almost completely the constraints, except for a very few models if $X=1$ is chosen; c) Set 1 with $\xi<0.004$ is the one that has the largest number of models that do not satisfy the pQCD constraints (e.g. $\sim3\%$ if $X=1$ {and $\sim 40\%$ if $X=4$}). For $X=1$ the absolute maximum mass of the Set 1 models  drops from $\sim2.75M_\odot$ to $\sim 2.5M_\odot$ for models that satisfy  pQCD, {and for $X=4$ it drops to $\sim 2.15M_\odot$}. }

In \cite{Traversi:2020aaa} the authors have performed a Bayesian inference analysis to constrain the EOS using as framework a RMF model similar to the one considered in the present study, taking, however, $\xi=0$ and $\Lambda_\omega=0$ and using only observations to constrain the parameters.  They have tested several different priors and the possibility of $\Lambda$-hyperon onset. They have generally obtained larger radii for a 1.4$M_\odot$ star, possibly because they take $\Lambda_\omega=0$. As a consequence, they also get quite large values of the symmetry energy slope at saturation, except when the saturation symmetry energy takes values below 20 MeV. Besides,  in \cite{Traversi:2020aaa}  smaller maximum masses were obtained. This property is connected to the nuclear effective mass in this kind of model. The most probable effective masses obtained are generally above 0.7 nucleon mass. As shown in \cite{Weissenborn:2011kb} in the model used  in \cite{Traversi:2020aaa}, the larger the effective mass the smaller the maximum mass configuration. In the model applied in our study, this correlation does not exist because of the presence of the $\omega^4$ term. {In the study \cite{Huang:2023grj}, the authors also take astrophysical observations as the constraining power of the Bayesian inference which takes as the underlying framework the same used in our study. In this study, the nuclear physics constraints are minimal and are mainly included in choosing a narrower prior that takes into account some nuclear physics prior knowledge. It is very interesting to see that observations favor a large $\xi$ parameter, and, as a consequence a speed of sound square of the order of 0.4 $c^2$ in the center of massive stars.}

{In the Supplemental material, we present a few  selected models for NSs with maximum mass 2.0, 2.2, 2.4, 2.6, and 2.75 M$_\odot$ (the extreme one), namely BMPF$\_$most$\_$HESS, BMPF220, BMPF240, BMPF260, and BMPF275, respectively. Its model parameters together with its NMP and NS properties are given, respectively, in Tables II and III of the Supplemental Material. }

\section*{ACKNOWLEDGMENTS} 
This work was partially supported by national funds from FCT (Fundação para a Ciência e a Tecnologia, I.P, Portugal) under Projects No. UIDP/\-04564/\-2020, No. UIDB/\-04564/\-2020 and 2022.06460.PTDC. 
MBA, one of the authors, would like to thank the FCT for its support through the Ph.D. grant number 2022.11685.BD. The authors acknowledge the Laboratory for Advanced Computing at the University of Coimbra for providing {HPC} resources that have contributed to the research results reported within this paper, URL: \hyperlink{https://www.uc.pt/lca}{https://www.uc.pt/lca}.

\section*{Data availability}
The final posterior of the model parameters, the equation of states, and the solutions for the star properties obtained {with all the sets} can be obtained from the link \href{https://zenodo.org/record/7854112}{\bf (10.5281/zenodo.7854111)}.

%
\onecolumngrid
\clearpage

\setcounter{figure}{0}
\setcounter{table}{0}
\section*{Supplemental Material}
{We make available the full posterior with 17829 model parameters, the corresponding equation of states, and their solutions for the star properties obtained for prior Set 0 (see the main article). Models are named chronologically as BMPF \{$x$\} 
with $x\in[0,17828]$.} {[(B)ayesian, Tuhin (M)alik, Constan\c ca (P)rovid\^encia, Márcio (F)erreira]}. 
{In addition, this material presents the median and its 90\% CI values of the RMF model parameters obtained with the prior Sets 0, 1, 2, and 3. (See the main article for details). We will also present a few of the selected models for NS maximum mass 2.0, 2.2, 2.4, 2.6, and 2.75 M$_\odot$ (the extreme one), namely BMPF$\_$most$\_$HESS, BMPF220, BMPF240, BMPF260, and BMPF275, respectively. In Set 1, the closest match from 1$\sigma$ (68 \% CI) to the HESS J1731-34 data is the BMPF$\_$most$\_$HESS, while others are selected from Set 0.}

\begin{table*}[h]
\caption{The median value along with 90\% CI "min" and "max" values of the RMF model parameter obtained with prior Set 0, 1, 2, and 3. The nucleon, $\omega$ meson, $\sigma$ meson, and $\varrho$ meson masses are 939, 782.5, 500, and 763 MeV, respectively.}
\label{tab:my-table}
\setlength{\tabcolsep}{4.5pt}
      \renewcommand{\arraystretch}{1.4}
\begin{tabular}{cccclccclccclccc}
\hline \hline 
\multirow{3}{*}{Parameter} & \multicolumn{3}{c}{Set 0}                             &  & \multicolumn{3}{c}{Set 1}                             &  & \multicolumn{3}{c}{Set 2}                             &  & \multicolumn{3}{c}{Set 3}                             \\ \cline{2-4} \cline{6-8} \cline{10-12} \cline{14-16} 
                           & \multirow{2}{*}{median} & \multicolumn{2}{c}{90\% CI} &  & \multirow{2}{*}{median} & \multicolumn{2}{c}{90\% CI} &  & \multirow{2}{*}{median} & \multicolumn{2}{c}{90\% CI} &  & \multirow{2}{*}{median} & \multicolumn{2}{c}{90\% CI} \\ \cline{3-4} \cline{7-8} \cline{11-12} \cline{15-16} 
                           &                         & min          & max          &  &                         & min          & min          &  &                         & min          & min          &  &                         & min          & max          \\ \hline
$g_\sigma$       & $8.454$  & $8.010$  & $9.691$  &  & $8.243$  & $7.952$  & $8.998$  &  & $8.683$  & $8.180$  & $9.657$  &  & $9.848$  & $9.088$  & $11.233$ \\
$g_\omega$       & $9.915$  & $9.084$  & $12.167$ &  & $9.458$  & $8.980$  & $10.956$ &  & $10.377$ & $9.490$  & $12.082$ &  & $12.466$ & $11.272$ & $14.613$ \\
$g_\rho$         & $12.193$ & $9.546$  & $14.599$ &  & $12.170$ & $9.481$  & $14.543$ &  & $12.204$ & $9.527$  & $14.580$ &  & $12.655$ & $9.847$  & $15.231$ \\
$b \times 10^3$  & $4.586$  & $2.205$  & $6.903$  &  & $5.420$  & $3.137$  & $7.359$  &  & $4.017$  & $2.204$  & $5.847$  &  & $2.205$  & $1.251$  & $3.559$  \\
$c \times 10^3$  & $-1.985$ & $-4.627$ & $3.530$  &  & $-1.839$ & $-4.638$ & $3.894$  &  & $-2.250$ & $-4.616$ & $2.960$  &  & $-1.146$ & $-3.272$ & $0.806$  \\
$\xi$            & $0.004$  & $0.000$  & $0.016$  &  & $0.002$  & $0.000$  & $0.004$  &  & $0.007$  & $0.004$  & $0.014$  &  & $0.018$  & $0.015$  & $0.027$  \\
$\Lambda_\omega$ & $0.064$  & $0.036$  & $0.103$  &  & $0.075$  & $0.041$  & $0.108$  &  & $0.057$  & $0.035$  & $0.088$  &  & $0.039$  & $0.030$  & $0.056$   \\ \hline
\end{tabular}
\end{table*}

\begin{table*}[h]
\caption{A few selected RMF model parameters from Set 0 for NS maximum mass 2.2 M$_\odot$, 2.2 M$_\odot$,2.4 M$_\odot$,2.6 M$_\odot$, and 2.75 M$_\odot$ (the extreme one). We also select a model from Set 1, BMPF$\_$most$\_$HESS: which is the closest to HESS posterior from 1 $\sigma$ interval among all others. The nucleon, $\omega$ meson, $\sigma$ meson, and $\varrho$ meson masses considered are 939, 782.5, 500, and 763 MeV, respectively.}
\setlength{\tabcolsep}{10.5pt}
      \renewcommand{\arraystretch}{1.4}
\label{tab:bmpf}
\begin{tabular}{cccccccc}
\hline \hline 
model            & $g_\sigma$ & $g_\omega$ & $g_\rho$ & $b \times 10^3$ & $c \times 10^3$ & $\xi$   & $\Lambda_\omega$ \\ \hline
BMPF\_most\_HESS	&	$8.113$	&	$9.189$	&	$9.968$	&	$6.336$	&	$-1.852$	&	$0.003$	&	$0.116$ \\
BMPF220	&	$8.516$	&	$10.193$	&	$11.261$	&	$3.378$	&	$0.643$	&	$0.002$	&	$0.051$  \\
BMPF240	&	$9.427$	&	$11.708$	&	$14.039$	&	$3.091$	&	$-3.484$	&	$0.002$	&	$0.039$   \\
BMPF260	&	$10.001$	&	$12.721$	&	$10.035$	&	$2.303$	&	$-2.836$	&	$0.002$	&	$0.038$   \\
BMPF275	&	$10.412$	&	$13.219$	&	$11.180$	&	$2.541$	&	$-3.586$	&	$0.001$	&	$0.028$ \\
 \hline
\end{tabular}
\end{table*}

\begin{figure}
\includegraphics[width=0.8\linewidth]{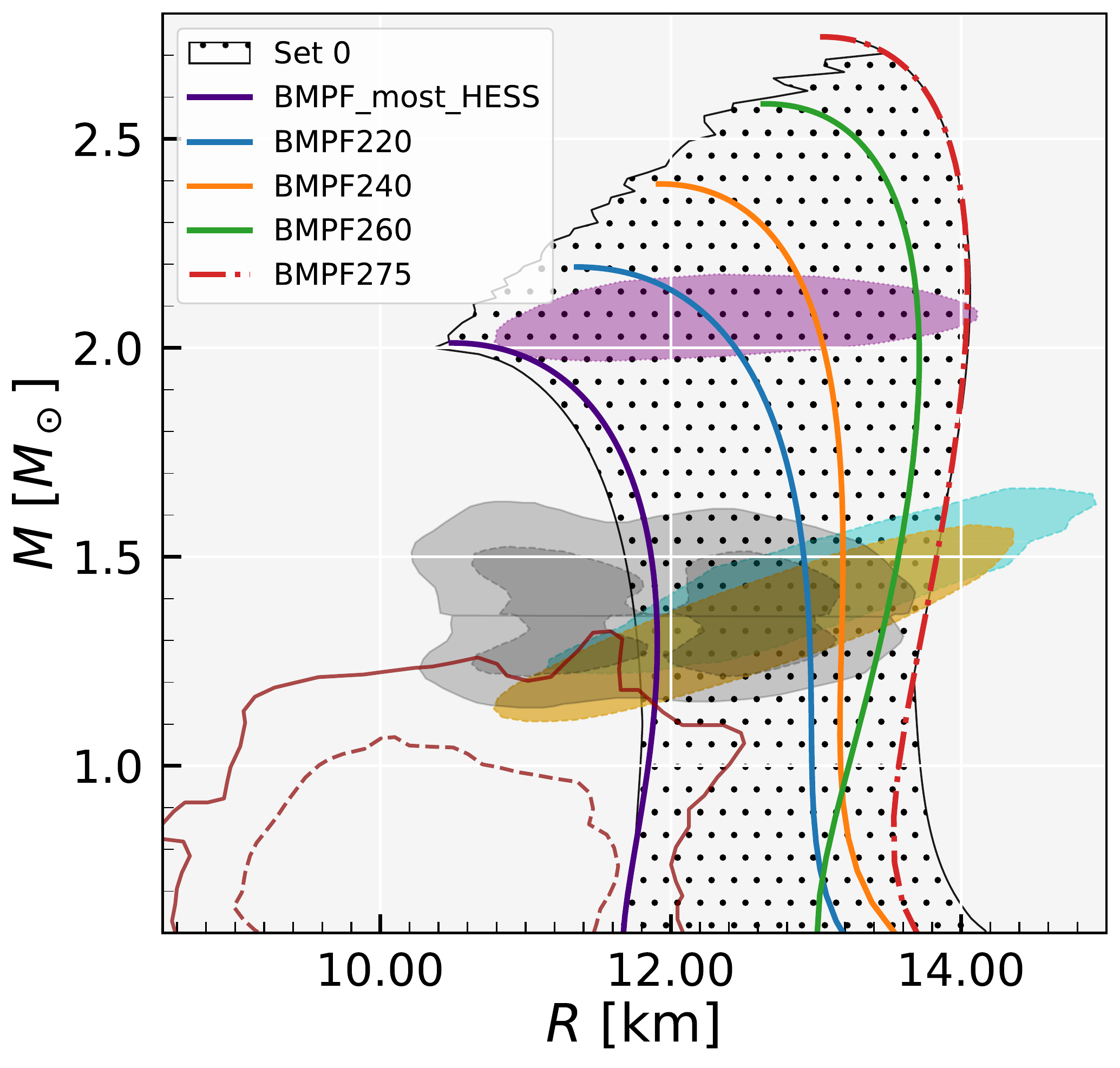}
\caption{{NS mass-radius domains 
(full posterior) produced in the following three scenarios: 
set 0 ($\xi\in[0,0.04]$).  The gray lines depict the constraints from the binary components of GW170817, along with their 90\% and 50\% credible intervals (CI).  The $1\sigma$ (68\%) CI for the 2D posterior distribution in the mass-radii domain for millisecond pulsar PSR J0030 + 0451 (cyan and yellow) 
\cite{Riley:2019yda, Miller:2019cac} as well as PSR J0740 + 6620 (violet) \cite{Riley:2021pdl, Miller:2021qha}  from the NICER x-ray data are also shown. Additionally, we show the constraint obtained from HESS J1731-347 for 68.3\% (95.4\%) CIs in dashed dark red (solid dark red) \cite{hess}.  MR curves for selected EOS displayed in table \ref{tab:bmpf}, resulted in NS maximum mass 2.0, 2.2, 2.4, 2.6, and 2.75 (the extreme one).}
\label{T:fig10}}
\end{figure}

\begin{table*}[]
\caption{The values of the NMPs and associated NS properties, the gravitational mass $M_{\rm max}$, baryonic mass $M_{\rm B, max}$, radius $R_{\rm max}$, central energy density $\varepsilon_ c$, central number density for baryon $\rho_c$, and square of central speed-of-sound $c_s^2$ of the maximum mass NS, the radius   $R_{{\rm M}_i}$ and  the dimensionless tidal deformability $\Lambda_{{\rm M}_i}$  ($\Lambda_{{\rm M}_i}$ for NS mass ${\rm M}_i \in [1.4,1.6,1.8,2.08]$ $M_\odot$), and the effective tidal deformability $\tilde \Lambda$ for the GW170817 merger with $q=1$ ($q$ is the mass ratio of NSs engaged in the binary merger) computed for those selected models displayed in table \ref{tab:bmpf}. }
\label{tab:prop}
\setlength{\tabcolsep}{8.5pt}
      \renewcommand{\arraystretch}{1.4}
\begin{tabular}{cccccccc}
\hline  \hline 
\multicolumn{2}{c}{\multirow{2}{*}{Quantity}}    & \multirow{2}{*}{Units} & \multicolumn{5}{c}{Selected Models}                          \\ \cline{4-8} 
\multicolumn{2}{c}{}                             &                        & BMPF\_most\_HESS & BMPF220  & BMPF240  & BMPF260  & BMPF275  \\ \hline
\multirow{11}{*}{NMP} & $\rho_0$                 & fm$^{-3}$              & $0.158$          & $0.146$  & $0.150$  & $0.149$  & $0.155$  \\
                      & $m^\star$                & …                      & $0.76$           & $0.74$   & $0.66$   & $0.61$   & $0.57$   \\
                      & $\varepsilon_0$          & \multirow{9}{*}{MeV}   & $-16.44$         & $-15.93$ & $-15.66$ & $-16.02$ & $-16.08$ \\
                      & $K_0$                    &                        & $258$            & $288$    & $234$    & $244$    & $177$    \\
                      & $Q_0$                    &                        & $-475$           & $-336$   & $-315$   & $-39$    & $-74$    \\
                      & $Z_0$                    &                        & $1852$           & $594$    & $4443$   & $8280$   & $18944$  \\
                      & $J_{\rm sym,0}$          &                        & $27.40$          & $31.74$  & $34.33$  & $29.06$  & $32.80$  \\
                      & $L_{\rm sym,0}$          &                        & $37$             & $45$     & $36$     & $56$     & $64$     \\
                      & $K_{\rm sym,0}$          &                        & $-105$           & $-159$   & $-34$    & $2$      & $77$     \\
                      & $Q_{\rm sym,0}$          &                        & $1082$           & $1215$   & $1649$   & $1132$   & $1741$   \\
                      & $Z_{\rm sym,0}$          &                        & $-7778$          & $-5469$  & $-18698$ & $-10622$ & $-17088$ \\
                      &                          &                        &                  &          &          &          &          \\
\multirow{15}{*}{NS}  & $M_{\rm max}$            & M $_\odot$             & $2.018$          & $2.200$  & $2.400$  & $2.592$  & $2.753$  \\
                      & $M_{\rm B, max}$         & M $_\odot$             & $2.394$          & $2.614$  & $2.894$  & $3.155$  & $3.387$  \\
                      & $c_{s}^2$                & $c^2$                  & $0.60$           & $0.62$   & $0.63$   & $0.63$   & $0.71$   \\
                      & $\rho_c$                 & fm$^{-3}$              & $1.122$          & $0.960$  & $0.845$  & $0.740$  & $0.683$  \\
                      & $\varepsilon_{c}$        & MeV fm$^{-3}$          & $1422$           & $1224$   & $1078$   & $942$    & $875$    \\
                      & $R_{\rm max}$            & \multirow{5}{*}{km}    & $10.49$          & $11.35$  & $11.92$  & $12.64$  & $13.03$  \\
                      & $R_{1.4}$                &                        & $11.90$          & $12.94$  & $13.18$  & $13.51$  & $13.78$  \\
                      & $R_{1.6}$                &                        & $11.80$          & $12.87$  & $13.18$  & $13.62$  & $13.89$  \\
                      & $R_{1.8}$                &                        & $11.57$          & $12.73$  & $13.15$  & $13.69$  & $13.97$  \\
                      & $R_{2.075}$              &                        & $...$            & $12.25$  & $12.99$  & $13.70$  & $14.04$  \\
                      & $\Lambda_{1.4}$          & \multirow{5}{*}{…}     & $366$            & $548$    & $581$    & $828$    & $844$    \\
                      & $\Lambda_{1.6}$          &                        & $139$            & $226$    & $257$    & $379$    & $397$    \\
                      & $\Lambda_{1.8}$          &                        & $53$             & $98$     & $124$    & $190$    & $206$    \\
                      & $\Lambda_{2.075}$        &                        & $...$            & $26$     & $44$     & $75$     & $86$     \\
                      & $\tilde \Lambda_{q=1.0}$ &                        & $435$            & $644$    & $676$    & $958$    & $973$    \\ \hline
\end{tabular}
\end{table*}

\end{document}